\documentclass[a4paper,12pt]{JHEP3}

\usepackage{amsmath,epsfig}
\usepackage{amssymb,amsfonts}
\usepackage{latexsym}
\usepackage[latin1]{inputenc}
\usepackage{xcolor}

\usepackage{graphicx}
\usepackage{longtable}

\relax
\renewcommand{\theequation}{\arabic{section}.\arabic{equation}}
\def\be{\begin{equation}}
\def\ee{\end{equation}}
\def\bea{\begin{eqnarray}}
\def\eea{\end{eqnarray}}

\newcommand\fverb{\setbox\pippobox=\hbox\bgroup\verb}
\newcommand\fverbdo{\egroup\medskip\noindent%
                        \fbox{\unhbox\pippobox}\ }
\newcommand\fverbit{\egroup\item[\fbox{\unhbox\pippobox}]}

\newcommand{\bear}{\begin{eqnarray}}

\newcommand{\eear}{\end{eqnarray}}

\newcommand{\bsea}{\begin{subeqnarray}}
\newcommand{\esea}{\end{subeqnarray}}
\newbox\pippobox

\def\6{\partial}

\def\a{\alpha}

\def\pa{\partial}

\def\m{\mu}

\def\sp{\;\;\;,\;\;\;}

\def\sq
\def\a{\alpha}

\def\hri#1#2{\href{http://arxiv.org/abs/#1}{[ArXiv:#1]#2}}
\def\hre#1#2{\href{http://arxiv.org/abs/#1/#2}{[ArXiv:#1/#2]}}

\allowdisplaybreaks[1]

\title{Hyperscaling-Violating Lifshitz hydrodynamics from black-holes: Part II}
\author{\Large   Elias Kiritsis$^{a,b,c}$, Yoshinori Matsuo$^{d}$\\
~\\
~\\
$^a$\href{http://hep.physics.uoc.gr}{Crete Center for Theoretical Physics, Institute of Theoretical and Computational Physics},
Department of Physics, University of Crete, 71003 Heraklion, Greece.
~\\
$^b$Crete Center for Quantum Complexity and Nanotechnology,
Department of Physics, University of Crete, 71003 Heraklion, Greece.
~\\
$^c$\href{http://www.apc.univ-paris7.fr/APC_CS/}{APC}, Univ Paris Diderot, Sorbonne Paris Cit\'e, UMR 7164 CNRS, F-75205 Paris, France.
~\\
$^d$%
Department of Physics, National Taiwan University,
Taipei 10617, Taiwan, R.O.C.
~\\
E-mail: \href{http://hep.physics.uoc.gr/~kiritsis/}{http://hep.physics.uoc.gr/~kiritsis/},
matsuo@phys.ntu.edu.tw
}

\preprint{CCTP-2016-7\\
CCQCN-2016-152}

\abstract{The derivation of Lifshitz-invariant hydrodynamics from holography, presented in \cite{km1} is generalized to arbitrary hyperscaling violating Lifshitz scaling theories with an unbroken U(1) symmetry.
The hydrodynamics emerging is non-relativistic with scalar ``forcing".
By a redefinition of the pressure it becomes standard non-relativistic hydrodynamics in the presence of a  chemical potential for the mass current.
 The hydrodynamics is compatible with the scaling theory of Lifshitz invariance with hyperscaling violation.
The bulk viscosity vanishes while the shear viscosity to entropy ratio is the same as in the relativistic case.
We also consider the dimensional reduction ansatz for the hydrodynamics and clarify the difference with previous results suggesting a non-vanishing bulk viscosity.
}


\begin{document}

\section{Introduction, results and outlook}\label{intro}

Holography \cite{Maldacena:1997re, Gubser:1998bc, Witten:1998qj, Witten:1998zw}
is a correspondence between quantum field theory and gravity. It
 provides a powerful tool to analyze strong-coupling theories by using
the dual gravity description.
At finite temperature and in the long-wavelength regime,
holography provides a correspondence between fluids and black holes.
This fluid/gravity correspondence was first studied
in the AdS/CFT correspondence
by \cite{Son:2002sd, Policastro:2002se, Policastro:2002tn, Baier:2007ix} in the linearized approximation and in \cite{Bhattacharyya:2008jc} by studying the fluctuations of the bulk black-hole horizon.

Holography has also been applied to geometries with
Lifshitz or Schr\"odinger symmetry mainly for application to condensed matter systems
\cite{Son:2008ye, Balasubramanian:2008dm, kachru, taylor, Guica:2010sw, cgkkm, gk1, gk2}\footnote{For a recent review see \cite{Marika}.}.
Many condensed matter systems have non-relativistic scale invariance
\cite{Eagles:1969zz, Nozieres:1985zz, O'Hara:2002zz, Regal:2004zza, Bartenstein:2004zza,Patel:2016ymd}, and
some of them have Lifshitz or Schr\"odinger symmetry
\cite{Hornreich:1975zz, Mehen:1999nd, Ardonne:2003wa}.
Moreover, the hydrodynamics of charge and energy in such systems may be
interesting as has been argued recently for the case of cold fermions at unitarity, \cite{fu}, other strongly correlated systems, \cite{zaa} and graphene, \cite{gra}.
Recent experiments in various materials that are strongly coupled as well as graphene have indicated that the hydrodynamics of electrons is observable and it exhibits non-trivial shear viscosity,
\cite{g1,g2,g3,g4,g5,g6}.

The fluid/gravity correspondences
in Schr\"odinger spacetimes has been  studied in \cite{Herzog:2008wg, Rangamani:2008gi,Bra},
where the dual description for non-relativistic fluid mechanics was given.
A proposal for hydrodynamics in Lifshitz invariant theories was  considered
in \cite{Hoyos:2013qna, Hoyos:2013cba} from the  effective field theory point of view.

In \cite{cgkkm, gk1, gk2}, all quantum critical holographic scaling geometries with a $U(1)$ symmetry respecting translation invariance and spatial rotation invariance were classified in terms of three scaling exponents.
Two of them $(z,\theta)$ appear in the metric while another exponent,
which is referred to as $\zeta$ in \cite{cgkkm, gk1, gk2},
appears in the profile of  the $U(1)$ gauge field.%
\footnote{This charge exponent controls the anomalous scaling of the charge density, even if it is conserved. It has also been introduced independently in \cite{hartong-obers} and was studied in more detail in \cite{G1} and \cite{karch}). The reason for the existence of anomalous charge exponent despite conservation is the RG running of the bulk coupling for charged degrees of freedom.}
The exponent $z$ is the Lifshitz (dynamical) scaling exponent, and
$\theta$ is the hyperscaling violation exponent, \cite{gk1,sh}.
Even though such theories have been studied intensively, many of their aspects are still unclear. In particular, hydrodynamics with Lifshitz scaling symmetry is not fully understood.

In non-relativist quantum field theories, the geometry that captures the symmetry and dynamics is the Newton Cartan geometry, \cite{Son}.
For Lifshitz space-times, the dual field theory is non-relativistic and
the source terms at the boundary
are related to the torsional Newton-Cartan theory
\cite{Christensen:2013lma, Christensen:2013rfa, BHR,Hartong:2014oma, Hartong:2014pma, Hartong:2015wxa}.
In particular, the role of torsional Newton-Cartan geometry in the boundary structure of bulk non-relativistic solutions as well as on the boundary symmetries was investigated in \cite{Hartong:2014oma, Hartong:2014pma, Hartong:2015wxa}. Alternatively the boundary structure can be analyzed using the Hamilton-Jacobi method, \cite{RS,CP}.

In \cite{km1}, the correspondence between fluids in the torsional Newton-Cartan theory and black holes in Lifshitz space-times has been analyzed.
Lifshitz space-times with unbroken $U(1)$ gauge symmetry
were considered, that were solutions of the Einstein-Maxwell-dilaton (EMD) theories with constant potential.
Although the geometry has Lifshitz scaling symmetry with dynamical exponent $z$, there was a non-trivial gauge field in the bulk that introduced non-trivial scaling exponents  in the charged sector.
The black-hole solution of that theory was considered, boosted  using Galilean boosts and then  all parameters of the solution including the velocities, were made $\vec x$-dependent.
 Using the technique  introduced in  \cite{Bhattacharyya:2008jc},
the bulk equations of motion were solved order by order in boundary derivatives and the (fluid) stress-energy tensor was computed and renormalized.
The results found were as follows:

\begin{itemize}

\item The stress-energy tensor was expressed in terms of the fluid variables:
velocity field $v^i$, energy density $\mathcal E$ and pressure $P$,
but it also contained the (particle number) density $n$ and
external source $\mathcal A_i$ associated to the $U(1)$ symmetry current.
It satisfied the scaling condition for Lifshitz invariant theories

$$z\mathcal E = (d-1)  P\;.$$

\item It was found that the stress tensor satisfied the conservation equations of the Newton-Cartan theory.

\item The role of the (unbroken) U(1) symmetry in this class of theories is important. It was found that it behaves very closely to the U(1) mass conservation symmetry in non-relativistic hydrodynamics.

\item The fluid equations found were non-relativistic in contrast to the
relativistic fluid analysis in \cite{Hoyos:2013qna, Hoyos:2013cba}.
Even though the continuity equation and energy conservation equation
agree with those in the ordinary non-relativistic fluids,
the Navier-Stokes equation was different from that in the ordinary non-relativistic fluids.

\item By redefining variables and allowing a (Milne-invariant) Newton potential in the sources, \cite{Hartong:2014oma}-\cite{Hartong:2015wxa} the fluid equations can be mapped to the standard non-relativistic fluid equations coupled to the torsional Newton-Cartan geometry in the presence of a Newtonian potential.

\item It was found that the form of the fluid equations is independent of the Lifshitz exponent $z$ as well as of the (non-trivial) conduction exponent.
It is only the constitutive relations (equation of state) that depend on these scaling exponents.

\item The entropy satisfies the local thermodynamic relation with the energy density and pressure.
The divergence of the entropy current is non-negative, compatible with  the second law.

\end{itemize}

The hydrodynamics analysis of \cite{km1} was generalized beyond the hydrodynamic limit in \cite{GJSV} using numerical techniques.

In this paper, we generalize the studies in \cite{km1} to those
with hyperscaling-violation in the associated holographic geometries.
In \cite{km1}, we have considered only the $\theta=0$ cases
in which hyperscaling-violation comes only from the gauge field.
Here, we introduce a non-trivial coupling between the dilaton and
cosmological constant in order to consider $\theta\neq 0$.
Fluids now, have hyperscaling-violation.
We find the following results:

\begin{itemize}
\item
As in \cite{km1}, the fluid equations and the stress-energy tensor reproduce those in the Newton-Cartan theory if
the holographic gauge field is identified to that that enters the Newton-Cartan theory.

\item
The fluid equations are similar to those in \cite{km1} but for  additional external forcing terms
which come from the coupling to the dilaton
and its external source term. Note however that in the Lifshitz case with $\theta=0$,
although the dilaton was non-trivial, no such terms appeared in the hydrodynamics. Their appearance is therefore tied to hyperscaling violation in the bulk geometry.

\item With a judicious choice of the pressure, these forcing terms can be redefined away. The hydrodynamic equations are now equivalent to the non-relativistic hydrodynamics equations with a conserved mass current, but with an additional chemical potential for the mass current, that is given in (\ref{FluidVariablesZ}). This chemical potential is not related to external sources like the Newtonian potential but is directly related to the fact that the theory violates  hyperscaling.

\item
The bulk viscosity is zero even in the presence of  hyperscaling-violation.

\item The hydrodynamic equations respect the full scaling symmetry of Lifshitz invariance with hyperscaling violation as detailed in appendix \ref{app:ScaleDim}.
\end{itemize}

The Lifshitz space-time with hyperscaling-violation can be also obtained
by dimensional reduction from higher-dimensional Lifshitz space-times without hyperscaling violation, \cite{gk1}.
In this case, the model has two dilatons. We derive also the associated hydrodynamics from the black holes. We find the following:
\begin{itemize}
\item
The bulk viscosity is non-zero if we consider the naive dimensional reduction,
i.e., with trivial background in the extra dimensions and fixed volume.

\item
When the internal metric on the extra dimensions, or equivalently,
the volume of the extra dimensions satisfies a specific relation to the U(1) charge, the bulk viscosity becomes zero. This reduction gives the lower dimensional hydrodynamics we derived earlier.

\item We also present the hydrodynamic ansatz which describes a Lifshitz-invariant fluid with hyperscaling violation on a non-trivial (conformally-flat   ) boundary metric.
The hydrodynamic equations in such a metric turn out to be simpler than on flat space.

\end{itemize}

The hydrodynamic equations we derive are general and are given by
 (\ref{LifContinuity})-(\ref{LifChargeCons}) and supplemented by (\ref{a}), with independent variables the temperature $T$, the (mass) charge density $n$ and velocities $v^i$.
All other variables like the energy $\mathcal E$, the pressure $P$ and the
chemical potential as well as the transport coefficients are determined in terms of $T,n$ by constitutive relations that are dynamics/theory specific.%
\footnote{
Here, $n$ plays the role of the density of the non-relativistic fluid. We use this (and not the associated chemical potential) as a fluid variable because of the important role it plays in the realization of the non-relativistic momenta. It is the same reason that the mass density is usually used
in non-relativistic fluid mechanics.
}
Finally there is the Lifshitz invariance Ward identity, namely (\ref{WardHV}).

The $z,\theta$ dependence does not appear explicitly in the hydrodynamic equations. To leading order, it appears only through the constituent relations that express the energy, pressure and (mass) chemical potential as functions of temperature and mass density. To next order the transport coefficients may also depend on $z$ and $\theta$. For this reason we expect that although the hydrodynamic equations we derived strictly apply to the duals of EMD theories we considered, their validity is universal.

This analysis completes the derivation of fluid dynamics for non-relativistic scale-invariant and U(1)-invariant theories with Lifshitz scaling and a violation of hyperscaling. Remains open the case of such theories without a U(1) symmetry (ie. a broken U(1) symmetry). The case of a perfect fluid was studied recently in \cite{HO}. It is highly probable that the hydrodynamics in this case will turn out to be non-relativistic hydrodynamics in the absence of a conserved ``mass" current.

 A further interesting question concerns the appropriate hydrodynamics for QFTs that are described by RG flows  that interpolate between relativistic and non-relativistic theories.
To motivate the answer to this question, we consider first non-Lorentz invariant (but rotationally invariant) flows between Lorentz invariant fixed points,%
\footnote{%
The fact that the speed off light can vary on branes was pointed out first in \cite{kk}.}
\cite{tw,gn}, but where the velocity of light in the IR is different for that in the UV. In such a case, the hydrodynamics of this theory, is quasi-relativistic, but with a speed of light that is {\em temperature dependent}. This context resembles more the proposal of \cite{Hoyos:2013qna, Hoyos:2013cba}.

The example above suggests that in a (Lorentz-violating) RG flow from a CFT (with a un unbroken U(1) symmetry that is used to drive the breaking of Lorentz invariance) to an IR non-relativistic scaling (rotational invariant) geometry at an arbitrary temperature, the hydrodynamics will be again of the relativistic form (but with general equation of state) and with a speed of light $c(T)$ that is again temperature dependent. In the IR, $c(T\to 0)=\infty$ and the hydrodynamics reduces to the one found here with the U(1) symmetry becoming the mass-related symmetry. This is the standard non-relativistic limit of the relativistic hydrodynamics while all thermodynamic functions and transport coefficients are smooth functions of $T$ (if no phase transition exists at finite $T$). Otherwise they follow the standard behavior at phase transitions.

More general breaking of Lorentz invariance during RG flow must involve higher form fields of tensors in the bulk, and the details of the RG flow become complicated. It is plausible that a RG-covariant definition of velocities and other state functions is necessary in order to define hydrodynamics globally in the full energy landscape.\footnote{A proposal was put forward in \cite{mukho} but we think the appropriate setup is a bit different.}

This paper is organized as follows.
In Section~\ref{sec:Model}, we introduce the model and
its Lifshitz solution with hyperscaling-violation.
In Section~\ref{sec:HydroAnsatz}, we introduce the hydrodynamic ansatz.
We consider the derivative expansion and solve the equations of motion to first order.
We calculate the stress-energy tensor in Section~\ref{sec:StressTensor},
and discuss the relation to  non-relativistic fluids in Section~\ref{sec:FluidEquation}.
In Section~\ref{sec:DimRed}, we consider the relation to the dimensional reduction from
higher-dimensional Lifshitz space-times without hyperscaling-violation.
In Section~\ref{sec:NaiveAnsatz}, we show that a simpler hydrodynamic ansatz
gives a fluid moving in a non-trivial but conformally flat metric.
In Appendix~\ref{app:NewtonCartan}, we review
the Newton-Cartan theory and its application to fluids.
In Appendix~\ref{app:ScaleDim}, we discuss the scaling dimensions
of the fluid variables and other relevant coefficients.

\section{Hyperscaling-violating solutions and black holes}\label{sec:Model}

We consider a holographic model with Lifshitz scaling symmetry and hyperscaling-violation.
The gravity dual is $(d+1)$-dimensional Einstein gravity with a Maxwell field $A_\mu$ and a dilaton $\phi$.
The action is given by
\begin{equation}
 S
 = \frac{1}{16\pi G} \int d^{d+1} x \sqrt{-g}
 \left(R - 2 \Lambda e^{- \nu \phi} - \frac{1}{2} (\partial\phi)^2 - \frac{1}{4} e^{\lambda\phi}F^2\right) \ ,
\label{BulkAction}
\end{equation}
where $F = dA$ is the field strength of the gauge field,
$\Lambda$ is a negative cosmological constant with a coupling to dilaton,
and $\lambda$ and $\nu$ are dimensionless coupling constants.
The equations of motion are given by
\begin{align}
 R_{\mu\nu}
 &=
 \frac{2 \Lambda}{d-1} e^{-\nu\phi} g_{\mu\nu}
 + \frac{1}{2} (\partial_\mu\phi)(\partial_\nu\phi)
 + \frac{1}{4} e^{\lambda\phi}
 \left(
  2 F_{\mu\rho} {F_\nu}^\rho - \frac{1}{d-1} F^2 g_{\mu\nu}
 \right)
\label{EinsteinEq}
\\
 0
 &=
 \nabla_\mu (e^{\lambda\phi} F^{\mu\nu}) ,
\label{MaxwellEq}
\\
 \Box \phi
 &=
 \frac{1}{4} \lambda e^{\lambda\phi} F^2 - 2 \nu \Lambda e^{-\nu\phi}\ .
\label{DilatonEOM}
\end{align}
This model has the Lifshitz geometry with hyperscaling-violation as a solution;
\begin{align}
 ds^2 &= e^{2\chi} d\tilde s^2 \ , \label{hvLifMetric}
\\
 d\tilde s^2 &= - r^{2z} dt^2 + \frac{dr^2}{r^2} + \sum_i r^2 (dx^i)^2 , \label{LifMetric}
\\
 e^\chi &= e^{\chi_0} r^{-\theta/(d-1)}
\end{align}
The solution for the gauge field and dilaton are given by
\begin{align}
 A_t &= a \sqrt{\mu}\, r^{z+d-1-\theta} \ , &
 e^{\lambda\phi} &= e^{\lambda\phi_0} r^{-2 d_1} \ .
 \label{PureA}
\end{align}
where
\begin{equation}
 d_1 = (d-1) - \frac{d-2}{d-1} \theta \ .
\end{equation}
Here, the cosmological constant is related to a length scale  of the solution, which we fixed to be 1.
$\Lambda$ is then given by
\begin{equation}
 \Lambda = - \frac{(z-d-1-\theta)(z+d-2-\theta)}{2} \ .
\end{equation}
The exponents $z$, $\theta$ and constants $\phi_0$, $\chi_0$, $\mu$
are determined by the coupling constants in the action $\lambda$, $\nu$,
and the free parameter $a$ of the solution by the following relations;
\begin{align}
 \lambda^2 &= 2 \frac{(d-1) d_1^2}{[(z-1)(d-1)-\theta][(d-1)-\theta]} \ ,\label{la} \\
 \nu &= \frac{\theta \lambda}{(d-1) d_1} \ , \label{nu}\\
 \phi_0 &= -\frac{2}{\lambda}\left(1+\frac{\theta}{(d-1)(d-1-\theta)}\right)\log a \ , \label{Phi0}\\
 \chi_0 &= -\frac{\theta}{(d-1)(d-1-\theta)}\log a \ , \label{x0} \\
 \mu &= \frac{2(z-1)}{z+d-1-\theta} \ . \label{Constmua}
\end{align}

To simplify further calculations, we rescale the gauge field such that the solution becomes
\begin{equation}
 A_t = a r^{z+d-1-\theta} \ .
\end{equation}
the original action (\ref{BulkAction}) becomes
\begin{equation}
 S
 = \frac{1}{16\pi G} \int d^{d+1} x \sqrt{-g}
 \left(R - 2 \Lambda e^{- \nu \phi} - \frac{1}{2} (\partial\phi)^2 - \frac{\mu}{4} e^{\lambda\phi}F^2\right) \ .
\end{equation}

The following black hole geometry is also a solution of this theory;
\begin{align}
 ds^2 &= e^{2\chi} d\tilde s^2
\\
 d\tilde s^2 &= - r^{2z} f(r) dt^2 + \frac{dr^2}{f(r) r^2} + \sum_i r^2 (dx^i)^2 ,
\\
 e^\chi &= e^{\chi_0} r^{-\theta/(d-1)}
\end{align}
where
\begin{equation}
 f = 1 - \frac{r_0^{z+d-1-\theta}}{r^{z+d-1-\theta}} \ .
\label{WarpF}
\end{equation}
The radius of the horizon is given by $r_0$ and the Hawking temperature of the black hole is
\begin{equation}
 T = \frac{z+d-1-\theta}{4\pi} r_0^z \ .
\label{HawkingT}
\end{equation}
The gauge field and dilaton take almost the same form as in the  the zero temperature solution
\begin{align}
 A_t &= a (r^{z+d-1-\theta} - r_0^{z+d-1-\theta}) , &
 e^{\lambda\phi} &= e^{\lambda\phi_0} r^{-2d_1} .
\end{align}
but the constant mode of $A_t$ is chosen such that
$A_t$ vanishes at the horizon for regularity.

\section{Solving the hydrodynamics ansatz}\label{sec:HydroAnsatz}

In this paper, we focus on the case of $d=4$.
In some parts however,  we may give results in arbitrary dimension $d$.
Extensions to other dimensions (namely $d=3$ and $d>4$) are expected to be straightforward.

We introduce the hydrodynamic ansatz, which describes the fluid mechanics
on the field theory side.
For the regularity at the future horizon,
we transform the black-hole solution to Eddington-Finkelstein coordinates;
\begin{equation}
 dt_+ = dt + \frac{dr}{r^{z+1}f} \ .
\end{equation}
Hereafter, we will always work in Eddington-Finkelstein coordinates
and $t$ will henceforth stand for the null coordinate $t_+$.
In these coordinates,
the black hole solution for the metric and gauge field are expressed as
\begin{align}
 ds^2
 &=
 r^{2\chi} d\tilde s^2
\\
 d\tilde s^2
 &=
 - r^{2z} f dt^2 + 2 r^{z-1} dt\,dr + r^2 (dx^i)^2 \ ,
\\
 A
 &=
 a\left( r^{z+3-\theta} - r_0^{z+3-\theta}\right) dt
 - a r^{2-\theta} dr \ ,
\end{align}
where we have taken the gauge such that $A_r$ vanishes in the original Fefferman-Graham coordinates.
Then, we perform the Galilean boost
\begin{align}
 t &\to t \ ,
&
 x^i & \to x^i - v^i t \ ,
\end{align}
where $v^i$ is a set of constant parameters of the Galilean boost (that we call velocity).
Note that this is a general coordinate transformation and for constant $v^i$ it provides a new class of solutions parametrized by $\vec v$. Moreover, we assume in this paper\footnote{The hydrodynamics  of $z=1$ with hyperscaling violation
was considered in \cite{Kanitscheider:2009as} obtaining such solutions by dimensional reduction from higher dimensional AdS geometries.}  that $z\not= 1$ and therefore such a coordinate transformation keeps the boundary sources%
\footnote{
They are defined as the most divergent parts of the metric and gauge field near the boundary.
}
invariant, as it should. It therefore provides a different state (more precisely ensemble)  of the same boundary theory.

The black hole geometry becomes
\begin{align}
 ds^2 &= e^{2\chi} d\tilde s^2 \ , \label{BoostedMetric}
\\
 d\tilde s^2 &= - (r^{2z} f - v^2 r^2) dt^2 + 2 r^{z-1} dt dr - 2 r^2 v^i dt\, dx^i + r^2 (dx^i)^2 \ ,
\\
 e^\chi &= e^{\chi_0} r^{-\theta/3}
\end{align}
and the solution for the gauge field and dilaton remains invariant;
\begin{align}
 A
 &=
 a\left( r^{z+3-\theta} - r_0^{z+3-\theta}\right) dt
 - a r^{2-\theta} dr \ ,
&
 e^{\lambda\phi}
 &=
 e^{\lambda\phi_0} r^{-2d_1} \ . \label{BoostedGaugeDilaton}
\end{align}
In order to describe the dynamics of the fluid,
we replace the free parameters $r_0$, $v^i$ and $a$ by slowly varying functions of the boundary coordinates ($\vec x,t$).
However, this procedure generates a non-trivial background metric at the boundary since
$a$ appears as an overall factor of the metric.
In order to obtain a flat space background at the boundary,
we absorb this overall factor by introducing the additional coordinate transformation
\begin{equation}
 x^\mu \to e^{-\chi_0} x^\mu \, \label{CoordRedef}
\end{equation}
with $\chi_0$ given in (\ref{x0}),
before replacing the parameters by functions.
Then, the metric becomes
\begin{align}
 ds^2 &= r^{-2\theta/3}
 \left[
 - (r^{2z} f - v^2 r^2) dt^2 + 2 e^{\chi_0} r dt dr - 2 r^2 v^i dt\, dx^i + r^2 (dx^i)^2 \ ,
 \right]
\label{BoostedBH}
\end{align}
and the gauge field is also rescaled as
\begin{equation}
 A
 =
 a^{1+\frac{\theta}{3(3-\theta)}}\left( r^{z+3-\theta} - r_0^{z+3-\theta}\right) dt
 - a r^{2-\theta} dr \ .
\end{equation}
Since we have rescaled the time coordinate,
the Hawking temperature is also rescaled as
\begin{equation}
 T = \frac{z+d-1-\theta}{4\pi} \, e^{-\chi_0} r_0^z
 = \frac{z+d-1-\theta}{4\pi} \, a^{\frac{\theta}{(d-1)(d-1-\theta)}} r_0^z \ .
 \label{HawkingTres}
\end{equation}

Now, we replace the parameters $r_0$, $v^i$ and $a$
by slowly varying function of the boundary coordinates $x^\mu$.
Moreover, we promote the constant part of $A_i$, which is usually gauged away
to $x^\mu$-dependent functions as was also done in \cite{km1}.
Then, the metric, gauge field and dilaton become
\begin{align}
 ds^2 &= r^{-2\theta/3}
 \left[
 - r^{2z} f dt^2 + 2 e^{\chi_0(x)} r dt\, dr + r^2 (dx^i-v^i(x) dt)^2
 \right]
\label{BGmetric}
\\
 f &= 1 - \frac{r_0^{z+3-\theta}(x)}{r^{z+3-\theta}} \ ,
\\
 A
 &=
 a^{1+\frac{\theta}{3(3-\theta)}}(x)\left( r^{z+3-\theta} - r_0^{z+3-\theta}(x) \right) dt
 - a(x) r^{2-\theta} dr + \mathcal A_\mu(x) dx^\mu \ ,\label{gauge}
\\
 e^{\lambda\phi}
 &=
 e^{\lambda\phi_0(x)} r^{-2d_1} \ ,  \label{BGdilaton}
\end{align}
where $x^\mu$-dependence of $\phi_0(x)$ and $\chi_0(x)$ comes from that of $a(x)$.
We have also introduced $\mathcal A_t(x)$ and  $\mathcal A_i(x)$, that
originate in the constant parts of $A_t$ and $A_i$, respectively, and which have been  now  replaced by functions of the boundary coordinates.
The relations (\ref{Phi0})-(\ref{Constmua}) are however preserved.
The above is no longer a solution of the equations of motion,
and we must introduce additional correction terms.

We consider the derivative expansion in $t,\vec x$ and calculate
the first order solution for the hydrodynamic ansatz.
We first expand \eqref{BGmetric}-\eqref{BGdilaton} at a point,
which we can take $x^\mu = 0$ without loss of generality.
Then, we assume that the derivatives are small since
$x^\mu$ dependence appears only through the slowly varying functions
$r_0(x)$, $v^i(x)$, $a(x)$, etc., and expand with respect to the derivatives $\partial_\mu$;
\begin{align}
 g_{\mu\nu}
 &=
 g_{\mu\nu}^{(0)} + g_{\mu\nu}^{(1)} + \cdots \ ,
\\
 A_\mu
 &=
 A_\mu^{(0)} + A_\mu^{(1)} + \cdots \ ,
\\
 \phi
 &=
 \phi^{(0)} + \phi^{(1)} + \cdots \ ,
\end{align}
where $g_{\mu\nu}^{(n)}$, etc.\ stands for $n$-th order terms in the derivative expansion.
At the leading order of the derivative expansion, the equations of motion
contain only the leading order terms $g_{\mu\nu}^{(0)}$, $A_\mu^{(0)}$ and $\phi^{(0)}$
which are equivalent to the solutions \eqref{BoostedMetric}-\eqref{BoostedGaugeDilaton}
before replacing the parameters by slowly varying functions, and hence, are satisfied.
At the next-to-leading order, only the linear order terms of the derivatives $\partial_\mu$ appear,
and the equations of motion are not satisfied due to these linear order terms.
Now, we introduce the correction terms to \eqref{BGmetric}-\eqref{BGdilaton} as
\begin{align}
 g_{\mu\nu}
 &=
 g_{\mu\nu}^{(0)} + g_{\mu\nu}^{(1)} + h^{(1)}_{\mu\nu} + \mathcal O(\partial^2) \ ,
\label{Gcorr}
\\
 A_\mu
 &=
 A_\mu^{(0)} + A_\mu^{(1)} + a_\mu^{(1)} + \mathcal O(\partial^2) \ ,
\label{Acorr}
\\
 \phi
 &=
 \phi^{(0)} + \phi^{(1)} + \varphi^{(1)} + \mathcal O(\partial^2) \ ,
\label{Pcorr}
\end{align}
where $h_{\mu\nu}^{(n)}$, $a_\mu^{(n)}$ and $\varphi^{(n)}$ are the correction terms
at the $n$-th order of the derivative expansion, which start from $n=1$ since
the equations of motion satisfied without the correction terms at leading order.
At the next-to-leading order, or equivalently, at the linear order of the derivative expansion,
the equations of motion give the inhomogeneous linear differential equations for
the correction terms, $h_{\mu\nu}^{(1)}$, $a_\mu^{(1)}$ and $\varphi^{(1)}$.
The inhomogeneity comes from the first order terms of \eqref{BGmetric}-\eqref{BGdilaton},
namely, $g_{\mu\nu}^{(1)}$, $A_\mu^{(1)}$ and $\phi^{(1)}$.

By solving the differential equations for the correction terms,
we obtain the following first order solution of the derivative expansion;
\begin{align}
 ds^2 &= r^{-2\theta/3}
 \biggl[
 - r^{2z} f dt^2 + 2 a^{-\frac{\theta}{3(3-\theta)}} r^{z-1} dt dr
 + r^2 (dx^i - v^i dt)^2
\notag\\&\qquad\qquad
 + \frac{2}{3-\theta} a^{-\frac{\theta}{3(3-\theta)}} r^z \partial_i v^i dt^2
 - r^2 F_1(r) \sigma_{ij} (dx^i - v^i dt) (dx^j - v^j dt)
\notag\\&\qquad\qquad
 + 2 \left(F_3(r) \partial_i r_0 + F_5(r) \partial_i a\right)dt (dx^i - v^i dt)
 \biggr]
\label{SolGeom}
\end{align}
where the first line is the original solution and the the rest are the corrections terms.  $\sigma_{ij}$ is the shear tensor
\begin{equation}
 \sigma_{ij} = \left(\partial_i v^j + \partial_j v^i\right)
 - \frac{2}{3} \partial_k v^k \delta_{ij} \ ,
\end{equation}
and the functions $F_i(r)$ are given by
\begin{align}
 F_1(r)
 &=
 a^{-\frac{\theta}{3(3-\theta)}}
 \int_\infty^r dr' \frac{r^{\prime\,3-\theta}-r_0^{3-\theta}}
{r'(r^{\prime\,z+3-\theta}-r_0^{z+3-\theta})}  \ ,
\\
 F_2(r)
 &=
 \left( 2(z-1) r^{z+3-\theta} - (z-5+\theta) r_0^{z+3-\theta}\right) \int_\infty^r dr'\, \widehat F_1(r'),
\\
 F_3(r)
 &=
 - 2(z-1) a^{-1-\frac{\theta}{3(3-\theta)}} \int_\infty^r \frac{dr'}{r^{\prime\,6-z+\theta}} F_2(r')
\\
 F_4(r)
 &=
 \left( 2(z-1) r^{z+3-\theta} - (z-5+\theta) r_0^{z+3-\theta}\right) \int_\infty^r dr'\, \widehat F_2 (r') ,
\\
 F_5(r)
 &=
 \int_\infty^r \frac{dr'}{r^{\prime\,6-z+\theta}}
 \left(- 2(z-1) a^{-1-\frac{\theta}{3(3-\theta)}} F_4(r')
 + \frac{\theta}{3(3-\theta)} a^{-1-\frac{\theta}{3(3-\theta)}} r^{\prime\,3-\theta}
 \right)
\\
 \widehat F_i(r) &= \frac{r^{7-2\theta}\widetilde F_i}{(r^{z+3-\theta}-r_0^{z+3-\theta})[2(z-1) r^{z+3-\theta} - (z-5+\theta) r_0^{z+3-\theta}]^2}
\\
 \widetilde F_1 (r) &= \frac{z+3-\theta}{2(z-1)} a \frac{r_0^{z-\theta}}{r^{5-\theta}}
 \Bigl(2 (z-1) (5-\theta)r^{z+3-\theta} r_0^2 - z(z+3-\theta) r^{5-\theta} r_0^z
\notag\\&\qquad\qquad\qquad\qquad\qquad
  + (z-5+\theta)(z-2) r_0^{z+3-\theta} \Bigr)
\\
 \widetilde F_2(r)
 &=  - \frac{\theta}{6 (3-\theta) (z-1)} r^{-\theta -5} r_0^{-2 \theta }
 \Bigl(-4 (z-1)^2 r_0^{2 \theta } r^{2 z+6}-r^{2 \theta } r_0^{2 z+6} (z-5+\theta)^2
\notag\\&\qquad\qquad\qquad
  +r^{\theta +5} (z+3-\theta)^2 r_0^{\theta +2 z+1}
   +4 (z-1) (z-5+\theta) r_0^{\theta +z+3} r^{\theta+z+3}\Bigr)
 \ .
\end{align}

The first order solution for the gauge field is
\begin{align}
 A &= a(x) \left[ a^{\frac{\theta}{3(3-\theta)}} \left(r^{z+3-\theta}  - r_0^{z+3-\theta}(x)\right)
 - \frac{1}{3-\theta}r^{3-\theta} \partial_i v^i(x)\right]dt
\notag\\&\quad
 - a(x) r^{2-\theta} dr + \mathcal A_\mu(x) dx^\mu
 + \left(F_2(r) \partial_i r_0 + F_4(r) \partial_i a \right) (dx^i - v^i dt) \ ,
\label{SolA}
\end{align}
and the dilaton has no correction term,
\begin{equation}
 e^{\lambda\phi} = e^{\lambda\phi_0(x)} r^{-2d_1} \ , \label{SolPhi}
\end{equation}

We find that, in order for \eqref{SolGeom}, \eqref{SolA} and \eqref{SolPhi} to be a  solution of the equations of motion,
the functions $r_0(x)$, $v^i(x)$, $a(x)$ and $\mathcal A_\mu(x)$ must satisfy the following constrains;
\begin{align}
 0 &= \partial_t a + v^i \partial_i a - a \partial_i v^i  ,
\label{ConstA}
\\
 0 &= \partial_t r_0 + v^i \partial_i r_0 + \frac{1}{3-\theta} r_0 \partial_i v^i  ,
\label{ConstT}
\\
 0 &= \mathcal F_{ti} + v^j \mathcal F_{ji}
 + \frac{z+3-\theta}{2 (z-1)} a^{1+\frac{\theta}{3(3-\theta)}} r_0^{z+3-\theta}
 \left(z \frac{\partial_i r_0}{r_0} + \frac{\theta}{3(3-\theta)} \frac{\partial_i a}{a} \right)
\label{ConstS}
\end{align}
where $\mathcal F = d\mathcal A$.

\section{Calculation of the stress tensor}\label{sec:StressTensor}

We will calculate now the boundary stress-energy tensor.
The asymptotically Lifshitz space-times have anisotropic behavior
in temporal and spatial directions, and hence
it is convenient to introduce the vielbeins in order to consider
their general asymptotic behavior.

The metric is expressed in terms of the vielbein $E^A$ as
\begin{equation}
 ds^2 = - (E^0)^2 + \delta_{ab} E^a E^b + (E^r)^2 \ .
\end{equation}
For the asymptotic Lifshitz space-time in Fefferman-Graham coordinates,
the vielbein can be expressed as \cite{Ross:2011gu}
\begin{align}
 E^r &= e^\chi \frac{dr}{r} \ , &
 E^0_\mu &= e^\chi r^z \tau_\mu(r,x^\mu) \ , &
 E^a_\mu &= e^\chi r \hat e^a_\mu(r,x^\mu) \ , \label{Edef}
\end{align}
where the one forms $\tau(r,x^\mu)$ and $\hat e^a(r,x^\mu)$ have
a finite and non-degenerate limit near the boundary,  $r\to\infty$,
and provide  the characteristic quantities of Newton Cartan geometry.
Eq.\ \eqref{Edef} does not however fix the $r\to\infty$ limit of $\tau(r,x^\mu)$ and $\hat e^a(r,x^\mu)$, uniquely.
For our solution \eqref{SolGeom}, we take the vielbein $E^A$ such that
$\tau(r,x^\mu)$ and $\hat e^a(r,x^\mu)$ behave in $r\to\infty$ as
\begin{align}
 \tau_\mu dx^\mu &= dt \ ,
&
 \hat e^a_\mu dx^\mu &= dx^a - v^a dt \ .
\label{vl}
\end{align}
Then, the induced metric at transverse surfaces, $dr=0$, is expressed for large $r$ as%
\begin{align}
 \gamma_{\mu\nu}
 &=
 r^{-2\theta/3} \left(- r^{2z} \tau_\mu \tau_\nu + r^2 \delta_{ab} \hat e^a_\mu \hat e^b_\nu\right) \ ,
\label{InducedMetric}
\\
 \gamma^{\mu\nu}
 &=
 r^{2\theta/3} \left(- r^{-2z} \hat v^\mu \hat v^\nu + r^{-2} \delta^{ab} e_a^\mu e_b^\nu\right) \ ,
\label{InducedMetric2}\end{align}
where $\hat v^\mu$ and $e_a^\mu$ are the inverse vielbeins, which take the following form\footnote{ Note that our notations is somewhat different from \cite{Hartong:2014oma}-\cite{Hartong:2015wxa}. The detailed notation and variables used in summarized in appendix  \ref{app:Notation}.}
as $r\to\infty$ for \eqref{SolGeom};
\begin{align}
 \hat v^\mu \nabla_\mu &= \nabla_t + v^i \nabla_i \ ,
&
 e_a^\mu \nabla_\mu &= \nabla_a \ .
\label{hat}\end{align}

Relations (\ref{InducedMetric}) and (\ref{InducedMetric2}) do not completely fix the $r$-dependent vielbeins $\tau_{\m}$ $\hat e^a_{\m}$ and $\hat v^{\m}$.
For example, the one form $\tau_\mu$ may also have subleading terms of order $\mathcal O(r^{2-2z})$,
which appear  at the same order as the leading terms of $\hat e^a_\mu$. This would change $\hat e^a_\mu$ and $\hat v^{\m}$ to leading order.
This ambiguity is equivalent to a Milne boost at the boundary.
We fix this ambiguity in the sequel in order to proceed. We will discuss the action of Milne boosts at the boundary data later.

To summarize, for our solution we take
\be
\tau=(1,\vec  0)\sp \hat v=(1,\vec v)\sp \hat e^a_{\m}=\left(\begin{matrix} -v^1&1&0&0\\
-v^2&0&1&0\\
-v^3&0&0&1\end{matrix}\right) \sp e_a^{\m}=\left(\begin{matrix} 0&0&0\\ 1&0&0\\
0&1&0\\
0&0&1\end{matrix}\right)
\label{holo}\ee
that describe the asymptotic Newton Cartan geometry in what we called in \cite{km1} ``the holographic frame".

For the gauge field with boundary  Lorentz indices we define
\begin{align}
 \hat A_0 &\equiv \hat v^\mu A_\mu=A_t+\vec v\cdot \vec A \ , &
 \hat A_a &\equiv e_a^\mu A_\mu \ ,
\label{Ahat}
\end{align}

The variation of the action with respect to these boundary variables is given by
\begin{equation}
 \delta S_r
 = \int d^4x \sqrt{-\gamma}\left(
 - \hat S^0_\mu \delta \hat v^\mu + \hat S^a_\mu \delta e_a^\mu
 + \hat J^0 \delta\hat A_0 + \hat J^a \delta\hat A_a
 + \mathcal O_\phi \delta \phi
 \right) \ ,
\label{VariationOfAction}
\end{equation}
where $S_r$ is the renormalized action with appropriate (boundary) counter terms.
Since the volume form behaves as
\begin{equation}
 \sqrt{-\gamma} \sim r^{z+3-4\theta/3} \ ,
\end{equation}
the terms at $\mathcal O(r^{-z-3+4\theta/3})$ in the expectation values
give the regular contributions.%
\footnote{
Even though $\hat A_0$ starts from $\mathcal O(r^{z+3-\theta})$,
the leading contribution to the effective action is
$\hat J^0 \mathcal A_0$ and hence,
the  regular term in $\hat J^0$ originates at the same order as the others, namely
$\mathcal O(r^{-z-3+4\theta/3})$.
}

The definition above  does not give the ordinary stress-energy tensor but
the one with contributions from the gauge field and current.
It is related to the ordinary Brown-York tensor as
\begin{equation}
 \hat S^0_\nu \hat v^\mu - \hat S^a_\nu e^\mu_a
 = T_\text{(BY)}{}^\mu{}_\nu + J^\mu A_\nu + T_\text{(ct)}{}^\mu{}_\nu
\end{equation}
where $T_\text{(BY)}^{\mu\nu}$ is the Brown-York tensor which is defined
in terms of the extrinsic curvature $K_{\mu\nu}$ as
\begin{equation}
 T_\text{(BY)}^{\mu\nu} = \frac{1}{8\pi G} \left(\gamma^{\mu\nu} K - K^{\mu\nu} \right) \ ,
\label{BYdef}
\end{equation}
and $T_\text{(ct)}^{\mu\nu}$ is the terms from the counter terms.
By using appropriate counter terms, the stress-energy tensor becomes finite at the boundary $r\to\infty$
and we define
\begin{align}
 \widetilde T^\mu{}_\nu
 \equiv \lim_{r\to\infty} r^{z+3-4\theta/3} \left(\hat S^0_\nu \hat v^\mu - \hat S^a_\nu e^\mu_a\right) \ ,
\label{TtildeDef}
\end{align}
and the current is given by
\begin{align}
 J^\mu \equiv \lim_{r\to\infty} r^{z+3-4\theta/3}
 \left(\hat J^0 \hat v^\mu + \hat J^a e^\mu_a\right) \ ,
\end{align}
As we will see later, the stress tensor in (\ref{TtildeDef})  is also different from the ordinary stress-energy tensor
on the boundary as it contains the contributions from the external gauge field.

Now, we calculate the renormalized stress-energy tensor from the first order solution \eqref{SolGeom}.
In order to regularize the stress-energy tensor and other expectation values,
we introduce the following counter terms;
\begin{equation}
 S_\text{ct} =
 \frac{1}{16\pi G} \int d^4 x \sqrt{-\gamma}
 \left[
 - (5+z - 2 \theta) e^{-\frac{1}{2}\nu\phi}+ \frac{z+3-\theta}{2} \, e^{(\lambda - \frac{1}{2}\nu)\phi} \gamma^{\mu\nu} A_\mu A_\nu
 \right] \ .
\label{CT}
\end{equation}

Although the last counterterm is apparently not gauge invariant,
the effective action is still invariant under boundary gauge transformations (up to boundary terms on the boundary).
To see this, the second term in \eqref{CT} is expanded near the boundary as
\begin{equation}
 \sqrt{-\gamma}\, e^{(\lambda-\frac{1}{2}\nu)\phi} \gamma^{\mu\nu} A_\mu A_\nu
 \propto
 a^\frac{\theta}{3(3-\theta)} r^{z+3-4\theta/3}
 - a^\frac{\theta}{3(3-\theta)} r_0^{z+3-\theta}
 + \frac{2}{a} \left(\mathcal A_t + v^i \mathcal A_i\right) \cdots \ .
\end{equation}
Near the boundary, $r\to\infty$,
 the  first term is leading and will cancel with the divergent part in the original action. The second and third terms
give finite contributions to the effective action. The ellipsis denotes terms that are vanishing as we take the cutoff surface  to the boundary.
We therefore observe that indeed the finite terms are not gauge invariant but are linear in the gauge field.

Under the boundary gauge transformation $\delta\mathcal A = d \Lambda$,
the finite term in \eqref{CT} transforms as
\begin{equation}
 \delta_\Lambda S_\text{ct}
 \propto \int d^4 x \left[\partial_t \frac{\Lambda}{a} + \partial_i \frac{\Lambda v^i}{a}\right] \ ,
\label{cu2}\end{equation}
where we used \eqref{ConstA}.
Therefore the transformed terms are surface terms and should vanish at infinity on the boundary. The coefficient of these terms  is indeed the conserved current (see equation (\ref{cu})) and (\ref{cu}) can be written as
\begin{equation}
 \delta_\Lambda S_\text{ct}
 \propto \int d^4 x J^{\m}\pa_{\m}\Lambda=\int d^4 x (\pa_{\m}J^{\m})\Lambda=0
\end{equation}
 Therefore if no charge is leaking to infinity on the boundary the effective action is gauge invariant.

By using the counter terms \eqref{CT}, the renormalized stress-energy tensor is calculated as
\begin{align}
 \widetilde T^0{}_0
 &=
 \frac{1}{8\pi G}
 \left(- \frac{3-\theta}{2} a^{\frac{\theta}{3(3-\theta)}}  r_0^{z+3-\theta}
 - \frac{z-1}{a} v^i \mathcal A_i \right)
 \ ,
\\
 \widetilde T^i{}_0
 &=
 \frac{1}{8\pi G}
 \biggl[-\frac{z+3-\theta}{2} a^{\frac{\theta}{3(3-\theta)}} r_0^{z+3-\theta} v^i
 + \frac{z-1}{a} v^i \mathcal A_t
 + \frac{1}{2} r_0^{3-\theta} \sigma_{ij} v^j
\notag\\&\quad\qquad\qquad\qquad
 + \frac{z(z+3-\theta)}{4(z-1)} r_0^{2z-\theta} \left( \partial_i r_0
 + \frac{\theta}{3 (3-\theta) z} \frac{r_0}{a} \partial_i a \right)
 \biggr]
 \ ,
\\
 \widetilde T^0{}_i
 &=
 \frac{1}{8\pi G}\frac{z-1}{a} \mathcal A_i
 \ ,
\\
 \widetilde T^i{}_j
 &=
 \frac{1}{8\pi G}
 \left[\frac{z}{2} a^{\frac{\theta}{3(3-\theta)}} r_0^{z+3-\theta} \delta_{ij}
 - \frac{1}{2} r_0^{3-\theta} \sigma_{ij}
 + \frac{z-1}{a} v^i \mathcal A_j
 - \frac{z-1}{a} \left(\mathcal A_t + v^k \mathcal A_k \right) \delta_{ij}\right]
 \ .
\end{align}
The current $J^\mu$ is obtained as
\be
 J^0 = \frac{z-1}{16\pi G a} \sp
 J^i = \frac{z-1}{16\pi G a} v^i \ .
\label{cu}\ee
The expectation value of the dual operator of the dilaton $\phi$
is also calculated as
\begin{align}
 \langle {\mathcal O}_\phi\rangle
 &= - \frac{1}{16 \pi G}
 \frac{9(z-1)+(3-\theta)\theta}{\sqrt{6(3-\theta)[3(z-1)-\theta]}}
 a^{\frac{\theta}{3(3-\theta)}} {r_0^{z+3-\theta}}
\notag\\&\quad
 + \frac{1}{16 \pi G}
 \frac{12(z-1)(6-\theta)}{\sqrt{6(3-\theta)[3(z-1)-\theta]}}
 \frac{1}{a} \left(\mathcal A_t + v^i\mathcal A_i \right)
\label{svev} \ .
\end{align}

We also calculate the entropy current that is given by, \cite{Bhattacharyya:2008xc},
\begin{equation}
 J_S^\mu = \frac{\sqrt{h}}{4G} \frac{n^\mu}{n^0} \ , \label{EntropyCurrent}
\end{equation}
where $\sqrt{h}$ is the volume form on the time-slice at the horizon and
$n^\mu$ is the normal vector to the horizon.
By using the first order solution \eqref{SolGeom}, the entropy current becomes
\begin{align}
 J_S^0 &= \frac{1}{4G} r_0^{3-\theta} \ ,
\\
 J_S^i &= \frac{1}{4G} r_0^{3-\theta} v^i
 - \frac{1}{8(z-1)G} a^{-\frac{\theta}{3(3-\theta)}} r_0^{z-\theta}
 \left(z \partial_i r_0 + \frac{\theta}{3(3-\theta)} \frac{r_0}{a} \partial_i a\right) \ .
\end{align}

\section{Hyperscaling-violating Lifshitz hydrodynamics}\label{sec:FluidEquation}

\subsection{Thermodynamics}

In order to consider the relation between the form of the stress-energy tensor \eqref{TtildeDef} and that for fluids,
we first calculate the thermodynamic functions.
The energy $E$, entropy $S$ and charge $N$ in volume  $V$ (in the $x^i$ directions)
of the Lifshitz black hole geometry are given by\footnote{In this subsection  we return temporarily to arbitrary dimension, $d$.}
\begin{align}
 E
 &=
 \mathcal E V = \frac{d-1-\theta}{16\pi G} a^\frac{\theta}{(d-1)(d-1-\theta)} r_0^{z+d-1-\theta} V \ ,
\\
 S
 &=
 s V = \frac{1}{4G} r_0^{d-1-\theta} V \ ,
\\
 N
 &=
 n V = \frac{z-1}{16\pi G a} V \ ,
\end{align}
where we have also defined $\mathcal E$, $s$ and $n$,
which are densities of energy, entropy and charge.
Here we have also taken into account the effect of the coordinate redefinition \eqref{CoordRedef},
and hence the energy has an additional factor of $a^\frac{\theta}{(d-1)(d-1-\theta)}$.

Now, we consider the first law of thermodynamics;
\begin{equation}
 d E
 =
 T dS - P dV
 + \mu dN \ ,
\end{equation}
where the variations with respect to entropy $S$, volume $V$ and charge $N$
give the temperature $T$, pressure $P$ and chemical potential $\mu$,
which are calculated as
\begin{align}
 T
 &=
 \left(\frac{\partial E}{\partial S}\right)_{V,N}
 = \frac{z+d-1-\theta}{4\pi} a^\frac{\theta}{(d-1)(d-1-\theta)} r_0^{z} \ ,
\\
 P &=
 - \left(\frac{\partial E}{\partial V}\right)_{S,N}
 = \frac{1}{16\pi G} \left(z-\frac{\theta}{d-1}\right)
 a^\frac{\theta}{(d-1)(d-1-\theta)} r_0^{z+d-1-\theta}  \ ,
\label{NewPressure}
\\
 \mu &= \left(\frac{\partial E}{\partial N}\right)_{S,V}
 = - \frac{\theta}{(d-1)(z-1)}
 a^{1 + \frac{\theta}{(d-1)(d-1-\theta)}} r_0^{z+d-1-\theta} \ .
\label{ChemicalPotential}
\end{align}
For $d=4$, the energy density, pressure, entropy density, charge density and
chemical potential are given by
\begin{align}
 \mathcal E &= \frac{3-\theta}{16\pi G} a^{\frac{\theta}{3(3-\theta)}} r_0^{z+3-\theta} \ , &
 P &= \frac{1}{16\pi G} \left(z-\frac{\theta}{d-1}\right)
 a^{\frac{\theta}{3(3-\theta)}} r_0^{z+3-\theta} \ ,
\notag\\
 s &= \frac{1}{4G} r_0^{3-\theta} \ , &
 n &= \frac{z-1}{16\pi G} a^{-1}\ ,
\notag\\
 \mu &= - \frac{\theta}{3(z-1)} a^{1 + \frac{\theta}{3(3-\theta)}} r_0^{z+3-\theta} \ .
\label{FluidVariablesZ}
\end{align}

Note that at $\theta=0$, the chemical potential $\mu$ vanishes.

\subsection{Relation to hydrodynamics in Newton-Cartan theory}

In this section we will rewrite the boundary stress tensor in terms of the Newton-Cartan geometry data along the lines of \cite{Hartong:2014oma}-\cite{Hartong:2015wxa} following the detailed formalism of \cite{km1}.%

In the Newton-Cartan theory, (that is explained in more detail in appendix \ref{app:NewtonCartan}),
the spacelike vielbein and inverse timelike vielbein  $\bar e^a_\mu$ and $\bar v^\mu$,
transform under the Milne boost.
In this paper, $\bar v^\mu$ and $\bar e^a_\mu$ indicate
the vielbeins in an arbitrary (Milne) frame.

There is a special frame, the ``holographic frame,'' introduced in \cite{km1}, where
the vielbeins are given by
\be
\bar v^\mu = \hat v^\mu\sp \bar e^a_\mu = \hat e^a_\mu\;.
\ee
where $\hat v^{\mu}$ and $\hat e^a_{\mu}$ are defined on the gravity side, in (\ref{hat}),
and $\hat v^\mu$ always satisfies in any frame
\be
 \hat v^\mu = u^\mu\equiv (1,\vec v)\;,
 \ee
where $u^\mu$ is the fluid velocity.
The timelike vielbein and inverse spacelike vielbein do not transform
under the Milne boost and hence are simply referred to as $\tau_\mu$ and $e^\mu_a$, respectively.%

The renormalized boundary stress-energy tensor, which we have calculated in the previous section,
can be expressed in the following form;
\begin{align}
 \widetilde T^\mu{}_\nu
 &=
 - \mathcal E \hat v^\mu \tau_\nu + (P - n\mu) \hat h^\mu{}_\nu
 - \kappa \tau_\nu h^{\mu\rho} \partial_\rho T
 - \eta \sigma_{ab} e_a^\mu \hat e^b_\nu
 + n \hat v^\mu \mathcal A_\nu - n \hat v^\rho \mathcal A_\rho \delta^\mu{}_\nu \ ,
\label{Ttilde1}
\end{align}
where the spatial metric is defined in terms of the vielbeins $e_a^\mu$ and $\hat e^a_\mu$;
\begin{align}
 h^{\mu\nu} &= e_a^\mu e_a^\nu \ ,
&
 \hat h^\mu{}_\nu &= e_a^\mu \hat e^a_\nu \ .
\end{align}
The energy density $\mathcal E$, pressure $P$, particle number density $n$
and chemical potential $\mu$ are given by \eqref{FluidVariablesZ}.
The transport coefficients like heat conductivity $\kappa$, shear viscosity $\eta$ and
bulk viscosity $\zeta$ are also read off (in $d=4$) as
\begin{align}
 \kappa &= \frac{1}{8(z-1)G} a^{-\frac{\theta}{3(3-\theta)}} r_0^{z+1-\theta} \ ,
&
 \eta &= \frac{1}{16\pi G} r_0^{3-\theta} \ ,
&
 \zeta &= 0 \ , \label{Transport}
\end{align}
where the bulk viscosity $\zeta$ is the coefficient of the expansion $\partial_i v^i$
in the stress-energy tensor \eqref{Ttilde1}.
Hence,
$\zeta=0$ can be deduced from the absence of $\partial_i v^i$
in \eqref{Ttilde1}.
The stress-energy tensor can also be expressed in terms of
the energy flow $\widetilde{\mathcal E}^\mu$, momentum density $\widetilde{\mathcal P}_\mu$,
stress tensor $\widetilde{\mathcal T}^\mu{}_\nu$ and current $J^\mu$, which are defined by
\begin{align}
 \widetilde{\mathcal E}^\mu
 &=
 - \widetilde T_0^\mu{}_\nu \hat v^\nu
\\
 \widetilde{\mathcal P}_\nu
 &=
 \widetilde T_0^\mu{}_\rho \tau_\mu \hat h^\rho{}_\nu
\\
 \widetilde{\mathcal T}^\mu{}_\nu
 &=
 \widetilde T_0^\rho{}_\sigma \hat h^\mu{}_\rho \hat h^\sigma{}_\nu
\end{align}
where
\begin{equation}
 \widetilde T_0^\mu{}_\nu = \widetilde T^\mu{}_\nu
 + n\mu \hat h^\mu{}_\nu
 - J^\mu \mathcal A_\nu + J^\rho \mathcal A_\rho \delta^\mu{}_\nu \ .
\end{equation}
Then, $\widetilde T^\mu{}_\nu$ is written as
\begin{align}
 \widetilde T^\mu{}_\nu
 &=
 - \widetilde{\mathcal E}^\mu \tau_\nu + \hat v^\mu \widetilde{\mathcal P}_\nu
 + \widetilde{\mathcal T}^\mu{}_\nu - n\mu \hat h^\mu{}_\nu
 + J^\mu \mathcal A_\nu - J^\rho \mathcal A_\rho \delta^\mu{}_\nu
 \ .
\label{Ttilde2}
\end{align}
Comparing \eqref{Ttilde2} with \eqref{Ttilde1}, we obtain
\begin{align}
 \widetilde{\mathcal E}^\mu
 &=
 \mathcal E \hat v^\mu - \kappa h^{\mu\rho} \partial_\rho T
\\
 \widetilde{\mathcal P}_\nu
 &=
 0
\\
 \widetilde{\mathcal T}^\mu{}_\nu
 &=
 P \hat h^\mu{}_\nu - \eta \sigma_{ab} e_a^\mu \hat e^b_\nu
\\
 J^\mu
 &=
 n \hat v^\mu\label{current}
\end{align}
The above stress-energy tensor can be identified with that in the Newton-Cartan theory\footnote{There can be ambiguities in the definition of the stress tensor. In relativistic CFTs they are known to affect higher-order transport coefficients, \cite{nakayama}. In non-relativistic scaling theories the situation is even more sensitive as they affect even the ideal hydrodynamics equations as we will see further. Moreover, as we discuss in section 6, they also affect the bulk viscosity.}.

In the Newton-Cartan theory, the geometry is described by
the timelike vielbein $\tau_\mu$, spatial (inverse) metric $h^{\mu\nu}$,
timelike inverse vielbein $\bar v^\mu$ and gauge field $B_\mu$.
Here, the spacelike vielbein, its inverse and spatial metric with lower indices are denoted as
$\bar e^a_\mu$, $e_a^\mu$ and $\bar h_{\mu\nu}$.
The energy current $\mathcal E^\mu$, stress tensor $\mathcal T_{\mu\nu}$,
momentum density $\mathcal P_\mu$ and mass current $\mathcal J^\mu$ are
the conserved quantities associated to $\tau_\mu$, $h^{\mu\nu}$,
$\bar v^\mu$ and $B_\mu$, respectively.
For a generic fluid in Eckart frame, they are given by \cite{Jensen:2014ama}
\begin{align}
 \mathcal E^\mu
 &=
 \mathcal E u^\mu + \frac{1}{2} \rho u^2 u^\mu - \kappa h^{\mu\nu} \partial_\nu T
 + h^{\mu\rho} u^\sigma \mathcal T_{\rho \sigma} \ ,
\label{NewtonE}
\\
 \mathcal P_\mu &= \rho u_\mu \ ,
\label{NewtonP}
\\
 \mathcal T_{\mu\nu}
 &=
 P \bar h_{\mu\nu} + \rho u_\mu u_\nu
 - \eta \sigma_{\rho \sigma} \bar P^\rho{}_\mu \bar P^\sigma{}_\nu
 - \zeta \vartheta \bar h_{\mu\nu} \ ,
\label{NewtonT}
\\
 \mathcal J^\mu
 &=
 \rho u^\mu \ ,
\label{Mcurrent}
\end{align}
where
\be
\bar P^\mu{}_\nu \equiv  e_a^\mu \bar e^a_\nu \sp \vartheta\equiv \pa_i v^i
\ee
$\bar P^\mu{}_\nu$ is the projector to the spatial directions and $\vartheta$ is the expansion of the fluid.

The fluid velocity $u^\mu$ satisfies the normalization condition $\tau_\mu u^\mu = 1$
and is given by $u^\mu = (1,u^i)$ for $\tau = dt$.

In the Newton-Cartan theory,
the generic stress-energy tensor (not only for the fluid above)
is defined from the energy current $\mathcal E^\mu$, stress tensor $\mathcal T_{\mu\nu}$,
and momentum density $\mathcal P_\mu$
as\footnote{We always denote stress-energy tensors
by the letter $T$ and their spatial projection (the stress tensor) by the letter $\mathcal T$.}
\begin{equation}
 \bar T^\mu{}_\nu
 = - \mathcal E^\mu \tau_\nu + \bar v^\mu \mathcal P_\nu
 + h^{\mu\rho} \mathcal T_{\rho\nu} \ .
\label{Tbar}
\end{equation}
The Newton-Cartan theory has a symmetry which is known as Milne boost;%
\footnote{%
When the geometry is torsion-free then gauge invariance and Milne boost invariance can be simultaneously present. However in more general cases they are incompatible, \cite{Hartong:2014oma}-\cite{Hartong:2015wxa}. In the case described in this paper the geometry is torsion-free.}

\begin{align}
 \bar v^\mu &\to \bar v'{}^\mu = \bar v^\mu + h^{\mu\nu} V_\nu \ , \\
{B} &\to { B}' =
 { B} + \bar P_\mu^\nu V_\nu dx^\mu
 - \frac{1}{2} h^{\mu\nu} V_\mu V_\nu \tau_\rho dx^\rho \ .
\end{align}
and we can introduce the Milne boost invariant combination for the gauge field 1-form as \cite{Jensen:2014ama,Hartong:2015wxa}
\begin{align}
 \widehat B = B + u_\mu dx^\mu
 - \frac{1}{2} u^2 \tau_\rho dx^\rho \ , \label{HtoNewton}
\end{align}
where
\begin{align}
 u_\mu &= \bar h_{\mu\nu} u^\nu \ ,
&
 u^2 &= \bar h_{\mu\nu} u^\mu u^\nu \ .
\end{align}
We also define a Milne boost invariant stress-energy tensor as
\begin{equation}
 T^\mu{}_\nu = \bar T^\mu{}_\nu + \mathcal J^\mu B_\nu - \mathcal J^\mu \widehat B_\nu
 = \bar T^\mu{}_\nu - \mathcal J^\mu \left(u_\nu - \frac{1}{2}\tau_\nu u^2\right) .
\label{TandTbar}
\end{equation}
We choose  $\bar v^\mu= u^\mu$ and we have
\begin{align}
 u_\mu &= \bar h_{\mu\nu} u^\nu = \bar h_{\mu\nu} \bar v^\mu = 0 \ ,
\\
 u^2 &= \bar h_{\mu\nu} u^\mu u^\nu = 0 \ ,
\end{align}
and then, \eqref{HtoNewton} and \eqref{TandTbar} give
\begin{align}
 B &= \widehat B \ ,
&
 T^\mu{}_\nu &= \bar T^\mu{}_\nu \ .
\end{align}
Therefore, the Milne boost invariants $\widehat B$ and $T^\mu{}_\nu$ are the same as
$B$ and $\bar T^\mu{}_\nu$ in the $\bar v^\mu = u^\mu$ frame.

We further define the Milne boost invariants for the energy flow
$\widehat{\mathcal E}^\mu$, momentum density $\widehat{\mathcal P}_\mu$,
stress tensor $\widehat{\mathcal T}^\mu{}_\nu$ from the Milne boost-invariant stress-energy tensor $T^\mu{}_\nu$ as
\begin{align}
 \widehat{\mathcal E}^\mu
 &=
 - T^\mu{}_\nu u^\nu \ ,
\\
 \widehat{\mathcal P}_\nu
 &=
 T^\mu{}_\rho \tau_\mu \bar P^\rho_\nu \ ,
\\
 \widehat{\mathcal T}^\mu{}_\nu
 &=
 T^\rho{}_\sigma \bar P^\mu_\rho \bar P^\sigma_\nu \ ,
\end{align}
and then, $T^\mu{}_\nu$ is written as
\begin{align}
 T^\mu{}_\nu
 &\equiv
 - \widehat{\mathcal E}^\mu \tau_\nu + u^\mu \widehat{\mathcal P}_\nu
 + \widehat{\mathcal T}^\mu{}_\nu \ .
\label{DefMinv}
\end{align}
{}From \eqref{NewtonE}-\eqref{NewtonT}, \eqref{Tbar}, \eqref{TandTbar} and \eqref{DefMinv},
the Milne boost invariants are expressed as
\begin{align}
 \widehat{\mathcal E}^\mu
 &=
 \mathcal E u^\mu - \kappa h^{\mu\nu} \partial_\nu T \ ,
\label{Ehat}
\\
 \widehat{\mathcal P}_\mu &= 0 \ ,\label{Phat}
\\
 \widehat{\mathcal T}_{\mu\nu}
 &=
 P \bar h_{\mu\nu}
 - \eta \sigma_{\rho \sigma} \bar P^\rho{}_\mu\bar P^\sigma{}_\nu
 - \zeta \theta \bar h_{\mu\nu} \ ,
\label{cThat}
\end{align}
and here $\bar h^{\mu\nu}$ and $\bar P^\mu{}_\nu$ are given by those in $\bar v^\mu = u^\mu$ frame.
Then, \eqref{NewtonE}-\eqref{NewtonT} take the same form to \eqref{Ehat}-\eqref{cThat}, respectively,
for $v^i = u^i$ and $\zeta=0$.
Therefore, we identify the energy flow $\widetilde{\mathcal E}^\mu$,
momentum density $\widetilde{\mathcal P}_\mu$ and stress tensor $\widetilde{\mathcal T}^\mu{}_\nu$,
which are calculated from the black hole geometry in the previous section,
with the Milne boost-invariants
$\widehat{\mathcal E}^\mu$, $\widehat{\mathcal P}_\mu$ and $\widehat{\mathcal T}^\mu{}_\nu$, respectively.
We also identify the gauge field $\mathcal A_\mu$,
which originates in the constant mode of the bulk gauge field $A_\mu$,
with the Milne boost-invariant combination $\widehat B_\mu$ as
\begin{equation}
 \mathcal A_\mu = m \widehat B_\mu - \mu \tau_\mu \ , \label{AtoB}
\end{equation}
where $m$ is the coupling constant for the gauge field $B_\mu$,
which is  equivalently in the non-relativistic language the mass per particle.
Then, the mass current $\mathcal J^\mu$ is related to the particle number current $J^\mu$ as
\begin{equation}
 \mathcal J^\mu = m J^\mu \ .
\end{equation}
It is then straightforward to verify that the stress-energy tensor $\widetilde T^\mu{}_\nu$,
which we calculated from the black hole geometry,
is related to the Milne boost-invariant stress-energy tensor $T^\mu{}_\nu$ in the Newton-Cartan theory as
\begin{equation}
 \widetilde T^\mu{}_\nu =
 T^\mu{}_\nu + \mathcal J^\mu \widehat B_\nu - \mathcal J^\rho \widehat B_\rho \delta^\mu{}_\nu \ .
\label{TandTtilde}
\end{equation}

The conservation law for these Milne boost-invariants is given by \cite{Jensen:2014aia}
\begin{align}
 D_\mu \widehat{\mathcal E}^\mu
 &=
 \hat v^\mu \widehat H_{\mu\nu} \mathcal J^\nu
 - \frac{1}{2}\left(h^{\mu\rho}D_\rho \hat v^\nu
 + h^{\nu\rho} D_\rho \hat v^\mu\right) \widehat{\mathcal T}_{\mu\nu} \ , \label{ConsInvE}
\\
 h^{\rho\mu} h^{\sigma\nu} D_\rho \widehat{\mathcal T}_{\mu\nu}
 &=
 h^{\sigma\nu}
 \left[
  \hat v^\mu D_\nu \widehat{\mathcal P}_\mu - D_\mu \left(\hat v^\mu \widehat{\mathcal P}_\nu\right)
  + \widehat H_{\mu\nu} \mathcal J^\mu
 \right] \ ,
 \label{ConsInvT}
\end{align}
where $\widehat H = d\widehat B$.
The covariant derivative $D_{\m}$ is defined as usual in the Newton-Cartan geometry (see appendix \ref{app:NewtonCartan}).

In terms of the Milne boost invariant stress-energy tensor,
the conservation law can be expressed as
\begin{equation}
 D_\mu T^\mu{}_\nu = J^\mu \mathcal F_{\mu\nu} = \mathcal J^\mu \widehat H_{\mu\nu} \ ,
\end{equation}
since $\widehat{\mathcal P}_\mu=0$.
More details are given in  Appendix~\ref{app:NewtonCartan}.

The renormalized stress-energy tensor $\widetilde T^\mu{}_\nu$ satisfies
the conservation equations which are equivalent to \eqref{ConsInvE} and \eqref{ConsInvT}.
These equations  come from the following components
of the bulk equations of motion;
\begin{align}
 n^\mu\gamma^{\nu\rho} R_{\mu\nu}
 &=
 8\pi G n^\mu\gamma^{\nu\rho} T_{\mu\nu}^\text{(bulk)} \label{ConstEin} \ ,
\\
 n_\nu \nabla_\mu (e^{\lambda\phi} F^{\mu\nu})
 &= 0 , \label{ConstMax}
\end{align}
where $n_\mu$ and $\gamma_{\mu\nu}$ are the normal vector and
the induced metric on the boundary, respectively.
In order to derive the conservation equations for the first order fluids,
we have to calculate the constraint equations (\ref{ConstEin}), (\ref{ConstMax})  to  second order in the derivative expansion.
The correction terms at each order do not contribute to
the constraint equations at that same order, and hence,
we do not need to solve the differential equation for the correction terms to second order.

The constraint equations (\ref{ConstEin}), (\ref{ConstMax}) can be expressed in terms of
the fluid variables $\mathcal E$, $P$, $n$ and $v^i$ defined in (\ref{NewtonE})-(\ref{Mcurrent}) as
\begin{align}
 0 &= \partial_t \mathcal E + v^i \partial_i \mathcal E + (\mathcal E + P) \partial_i v^i
\notag\\&\qquad
 - \frac{1}{2} \eta \sigma_{ij}\sigma_{ij}
 - \partial_i( \kappa \partial_i T) \ ,
\label{LifContinuity}
\\
 0 &= \partial_i P - n \partial_i \mu + n \mathcal F_{ti} + n v^j \mathcal F_{ji}
  - \partial_j \left(\eta\sigma_{ij} \right) \ ,
\label{LifNavierStokes}
\\
 0 &= \partial_t n + \partial_j (n v^j) \ .
\label{LifChargeCons}
\end{align}
Note that the main difference in these equations for the hyperscaling violating case studied here and the case with $\theta=0$ studied in \cite{km1} is the appearance of the term $n\pa_i\mu$ in the Navier-Stokes-like equation (\ref{LifNavierStokes}). Here we have a chemical potential in the absence of an external Newtonian potential and this is due to hyperscaling violation.

The fluid equations \eqref{LifContinuity}-\eqref{LifChargeCons} are equations for
energy density $\mathcal E$, pressure $P$, velocity field $v^i$,
particle number density $n$ and temperature $T$.
There are also constituent relations between these variables.
In particular, the temperature and energy density are related to each other, and hence are not independent.
For the fluid dual to the Lifshitz geometry,
the energy density is constrained first by the Lifshitz Ward identity with hyperscaling violation. This is   given by
\begin{equation}
 \left(z-\frac{\theta}{d-1}\right) \mathcal E = (d-1-\theta) P \ .
\label{WardHV}
\end{equation}

The transport coefficients $\eta$ and $\kappa$ are not constant
but depend on the temperature, and also on the particle number density in this case, as in (\ref{Transport}).
By using \eqref{HtoNewton} and \eqref{AtoB},
the gauge field $\mathcal A$ is rewritten in terms of the velocity field $v^i$
and the external gauge field of $B$;
\begin{align}
 \mathcal A &=  m \left(B + u_\mu dx^\mu
 - \frac{1}{2} u^2 \tau_\rho dx^\rho \right) - \mu \tau
\notag\\
 &= m \left(B + v^idx^i
 - \frac{1}{2} \vec v^2  dt \right) - \mu dt .
\label{a}
\end{align}
Therefore, we have 5 equations for 5 independent variables $\mathcal E$, $v^i$ and $n$.

We also have external sources in the hydrodynamic equations:
one is of course the space metric that is flat here (but in subsequent sections this will change).
The other is the external gauge field $B_{\mu}$ that couples to the $U(1)$ current.
In ordinary non-relativistic hydrodynamics, $B_t$ would be the ordinary Newtonian potential
that couples to mass \cite{Hartong:2014oma,Hartong:2014pma,Hartong:2015wxa,km1}.

\subsection{The entropy current}

The entropy current \eqref{EntropyCurrent} can be expressed as
\begin{equation}
 J_S^\mu = s \hat v^\mu - \frac{\kappa}{T} h^{\mu\rho} \partial_\rho T
\end{equation}
where the entropy density $s$ is given by
\begin{equation}
 s = \frac{1}{4G} r_0^{3-\theta} \ .
\end{equation}
The entropy current satisfies the following relation with
the (internal) energy density $\widehat{\mathcal E}$ and pressure $P$;
\begin{align}
 T J_S^\mu = \widehat{\mathcal{E}}^\mu + P \hat v^\mu - \mu J^\mu
 = - T^\mu{}_\nu \hat v^\nu\ + (P - \mu n) \hat v^\mu  \ .
\label{ThermoRel}
\end{align}
It also satisfies the second law.
By using \eqref{LifContinuity}, the divergence of $J_S^\mu$ is expressed as
\begin{equation}
 \partial_\mu J_S^\mu
 = \frac{1}{2} \frac{\eta}{T} \sigma_{ij} \sigma_{ij} + \frac{\kappa}{T^2} \left(\partial_i T\right)^2 \ ,
\end{equation}
which is manifestly non-negative. Therefore, the entropy current satisfies the second law.
The entropy density $s = J_S^0$ is such that
the KSS bound, \cite{Kovtun:2004de}, is saturated;
\begin{equation}
 \frac{\eta}{s} = \frac{1}{4\pi} \ .
\end{equation}

\section{The relation to dimensional reduction}\label{sec:DimRed}

In the previous section, we studied the fluids in the field theory dual of
the Lifshitz space-time with hyperscaling-violation.
We found that the fluid has zero bulk viscosity $\zeta=0$.

It is known that the Lifshitz space-time with hyperscaling-violation
can be obtained by the dimensional reduction from
the higher-dimensional Lifshitz space-time without hyperscaling-violation, \cite{gk1}.
As discussed in \cite{km1}, the fluid for the Lifshitz space-time
without hyperscaling-violation also has zero bulk viscosity.
Upon compactification, this  Lifshitz geometry with flat internal space and constant radius, becomes the lower dimensional geometry with hyperscaling violation.
In such a case it was shown, \cite{Kanitscheider:2009as,gs}, that
the bulk viscosity becomes non-zero after  dimensional reduction.

In this section, we will first derive the general fluid equations in the higher-dimensional theory by allowing also the internal volume to be an additional thermodynamic variable.
Then we will reduce to the lower dimension. We will show that the appropriate reduction that corresponds to the thermodynamic ansatz in the lower dimension is compatible with our four-dimensional results of the previous section.

We first consider the following theory, which is
the Einstein gravity with the Maxwell field and a single scalar field;
\begin{equation}
 S
 = \frac{1}{16\pi G} \int d^{D+1} x \sqrt{-g_D}
 \left(R - 2 \Lambda - \frac{1}{2} (\partial\phi_1)^2 - \frac{1}{4} e^{\tilde\lambda\phi_1}F^2\right) \ ,
\label{ActionHigherDim}
\end{equation}
where $D=d-\theta$, and we assumed that $\theta$ is a negative integer at this moment.
This model has the Lifshitz space-time without hyperscaling-violation as a solution;
\begin{equation}
 ds^2_D = - r^{2z} dt^2 + \frac{dr^2}{r^2} + \sum_{i=1}^{d-\theta-1} r^2 (dx^i)^2 ,
\label{MetricHigherDim}
\end{equation}
with the following gauge field and dilaton;
\begin{align}
 A_t &= a \sqrt{\mu} \, r^{z+d-\theta-1} \ , &
 e^{\tilde\lambda\phi_1} &= a^{-2} r^{-2(d-\theta-1)} \ ,
\end{align}
where the parameters $z$, $a$ and $\mu$ are related to the parameters of the action (coupling constants) as
\begin{align}
 \tilde\lambda^2 &= 2 \frac{d-\theta-1}{z-1} \ , \\
 \Lambda &= - \frac{(z+d-\theta-1)(z+d-\theta-2)}{2} \ , \\
 \mu &= \frac{2(z-1)}{z+d-\theta-1} \ .
\end{align}

We compactify the $(-\theta)$-dimensional extra dimensions and
consider the dimensional reduction to $(d+1)$-dimensional spacetime.
The metric is decomposed as
\begin{align}
 ds^2_D &= e^{-\tilde\nu\phi_2}ds^2 + \sum_{i=d}^{d-\theta-1} e^{-(d-1)\tilde\nu\phi_2/\theta} (dx^i)^2 \ ,
\label{MetricHD}
\end{align}
where $ds^2$ is the metric after the dimensional reduction and
we compactified $x^i$-directions with $i=d,\cdots,d-\theta-1$
to circles with period $x^i \sim x^i + 1$.
The constant $\tilde\nu$ is given by
\begin{equation}
 \tilde\nu^2 = - \frac{2\theta}{(d-1)(d-\theta-1)} \ .
\end{equation}
Then, after the dimensional reduction, the action becomes
\begin{equation}
 S
 = \frac{1}{16\pi G} \int d^{d+1} x \sqrt{-g}
 \left(R - 2 \Lambda e^{- \tilde\nu \phi_2} - \frac{1}{2} (\partial\phi_1)^2
 - \frac{1}{2} (\partial\phi_2)^2
 - \frac{1}{4} e^{\tilde\lambda\phi_1 + \tilde\nu \phi_2}F^2\right) \ ,
\label{BulkActionDimRed}
\end{equation}
Now, the geometry has hyperscaling-violation because of
the redefinition of the metric, and is given by
\begin{align}
 ds^2 &= r^{-2\theta/(d-1)} d\tilde s^2
\\
 d\tilde s^2 &= - r^{2z} dt^2 + \frac{dr^2}{r^2} + \sum_{i=1}^{d-1} r^2 (dx^i)^2 .
\end{align}
while the additional dilaton field $\phi_2$ is
\begin{equation}
 e^{\tilde\nu\phi_2} = r^{-2\theta/(d-1)} \ .
\end{equation}
More generally, $\phi_2$ also has the constant mode as is $\phi_1$
but it is independent from the gauge field.
Then, the solution becomes
\begin{align}
 ds^2 &= e^{2\chi} d\tilde s^2
\\
 d\tilde s^2 &= - r^{2z} dt^2 + \frac{dr^2}{r^2} + \sum_{i=1}^{d-1} r^2 (dx^i)^2 , \label{gtilde}
\\
 e^{\tilde\nu\phi_2} &= e^{2\chi} = e^{2\chi_0} r^{-2\theta/(d-1)} \ ,
\end{align}
where $\chi_0$ is an arbitrary constant. We define a new parameter $b$ as
\begin{align}
 \chi_0 &= - \frac{\theta}{(d-1)(d-1-\theta)}\log b \ .
\end{align}
In a sense $b$ parametrizes the internal volume.

Now, the model has 2 dilatons instead of one.
This solution has an additional parameter $b$
which is the constant mode of the additional dilaton $\phi_2$.
The solution becomes equivalent to that in the previous sections
if the dilatons satisfy the following condition;
\begin{equation}
 \lambda\phi
 = \left(1 + \frac{\theta}{(d-1)(d-\theta-1)}\right) \tilde \lambda\phi_1
 = \left(1 + \frac{(d-1)(d-\theta-1)}{\theta} \right) \tilde\nu\phi_2 \ .
\end{equation}
In fact the above solution satisfies this condition
except for the constant mode $b$,
and hence, it is equivalent to the solution in the previous sections if
\begin{equation}
 b=a \ . \label{b=a}
\end{equation}
As already noted,  the value of $b$ controls the volume of the internal dimensions.
For the higher-dimensional Lifshitz geometry \eqref{MetricHigherDim},
we have
\begin{equation}
 b=1 \ ,
\end{equation}
but arbitrary $b$ can be introduced by the redefinition of
the coordinates in the extra dimensions.

The solution which describes the hydrodynamics
can be calculated in a similar fashion to the previous sections.
We now specialize to $d=4$.
The first order solution of the derivative expansion
is obtained as
\begin{align}
 ds^2_D
 &= d\tilde s^2 + \sum_{i=3}^{3-\theta} e^{-(d-1)\tilde\nu\phi_2/\theta} (dx^i)^2 \ ,
\label{SolBeforeDR}
\\
 ds^2 &= r^{-2\theta/(d-1)} e^{\nu\varphi_2} d\tilde s^2 \ ,
\label{SolAfterDR}
\\
 d\tilde s^2 &= - r^{2z} f dt^2 + 2 b^{-\frac{\theta}{3(3-\theta)}} r^{z-1} dt\,dr
 + r^2 (dx^i - v^i dt)^2
\notag\\&\quad
 + \frac{2}{3-\theta} b^{-\frac{\theta}{3(3-\theta)}} r^z \partial_i v^i dt^2
 - r^2 F_1(r) \sigma_{ij} (dx^i - v^i dt) (dx^j - v^j dt)
\notag\\&\quad
 - \frac{2\theta}{3(3-\theta)} F_1(r)
 \left[\partial_i v^i - b^{-1}\left(\partial_t b + v^i \partial_i b\right)\right]
 \left( - r^{2z} f dt^2 + 2 b^{-\frac{\theta}{3(3-\theta)}} r^{z-1} dt dr \right)
\notag\\&\quad
 + 2 \left(F_3(r) \partial_i r_0 + F_5(r) \partial_i b\right)dt (dx^i - v^i dt) \ ,
\\
 \varphi_2 &= \sqrt{-\frac{2\theta}{3(3-\theta)}} F_1(r)
 \left[\partial_i v^i - b^{-1}\left(\partial_t b + v^i \partial_i b\right)\right] \ .
\end{align}
where $\theta$ must be negative integer for the metric before the dimensional reduction $ds_D^2$.
The non-zero components of $v^i$ and $\mathcal A_i$ are introduced only for $i=1,2,3$ and
the parameters $v^i$, $r_0$, $a$, $b$ and $\mathcal A_\mu$ are replaced by functions
which depend on $x^\mu$ with $\mu = 0,\cdots,3$.
We have redefined the coordinates $x^\mu$ as \eqref{CoordRedef} in the previous section
and hence the Hawking temperature is also rescaled as
\begin{equation}
 T = \frac{z+d-1-\theta}{4\pi} \, e^{-\chi_0} r_0^z
 = \frac{z+d-1-\theta}{4\pi} \, b^{\frac{\theta}{(d-1)(d-1-\theta)}} r_0^z \ .
\label{HawkingTdr}
\end{equation}
where we have written this relation in arbitrary dimension.
The function $F(r)$ is given by
\begin{align}
 F_1(r)
 &=
 b^{-\frac{\theta}{3(3-\theta)}}
 \int dr \frac{r^{3-\theta}-r_0^{3-\theta}}{r(r^{z+3-\theta}-r_0^{z+3-\theta})}  \ ,
\\
 F_2(r)
 &=
 \left( 2(z-1) r^{z+3-\theta} - (z-5+\theta) r_0^{z+3-\theta}\right) \int dr\, \widehat F_1 (r)
\\
 F_3(r)
 &=
 - 2(z-1) a^{-1}b^{-\frac{\theta}{3(3-\theta)}} \int \frac{dr}{r^{6-z+\theta}} F_2(r)
\\
 F_4(r)
 &=
 \left( 2(z-1) r^{z+3-\theta} - (z-5+\theta) r_0^{z+3-\theta}\right) \int dr\, \widehat F_2 (r)
\\
 F_5(r)
 &=
 \int \frac{dr}{r^{6-z+\theta}}
 \left(- 2(z-1) a^{-1} b^{-\frac{\theta}{3(3-\theta)}} F_4(r)
 + \frac{\theta}{3(3-\theta)} b^{-1-\frac{\theta}{3(3-\theta)}} r^{3-\theta}
 \right)
\\
 \widehat F_i(r)
 &=
 \frac{r^{7-2\theta} \widetilde F_i(r)}{(r^{z+3-\theta}-r_0^{z+3-\theta})[2(z-1) r^{z+3-\theta} - (z-5+\theta) r_0^{z+3-\theta}]^2}
\\
 \widetilde F_1(r)
 &=
 \frac{z+3-\theta}{2(z-1)} a \frac{r_0^{z-\theta}}{r^{5-\theta}}
 \Bigl(2 (z-1) (5-\theta)r^{z+3-\theta} r_0^2 - z(z+3-\theta) r^{5-\theta} r_0^z
\notag\\&\qquad\qquad\qquad\qquad\qquad
  + (z-5+\theta)(z-2) r_0^{z+3-\theta} \Bigr)
\\
 \widetilde F_2(r)
 &=
 - \frac{\theta}{6 (3-\theta) (z-1)} a b^{-1} r^{-\theta -5} r_0^{-2 \theta }
 \Bigl(-4 (z-1)^2 r_0^{2 \theta } r^{2 z+6}-r^{2 \theta } r_0^{2 z+6} (z-5+\theta)^2
\notag\\&\qquad\qquad\qquad
  +r^{\theta +5} (z+3-\theta)^2 r_0^{\theta +2 z+1}
   +4 (z-1) (z-5+\theta) r_0^{\theta +z+3} r^{\theta+z+3}\Bigr) \ .
\end{align}

The first order solution of the gauge field is
\begin{align}
 \frac{1}{\sqrt\mu} A &= a(x) \left[b^{\frac{\theta}{3(3-\theta)}}\left(r^{z+3-\theta} - r_0^{z+3-\theta}(x)\right)
 - \frac{1}{3-\theta}r^{3-\theta} \partial_i v^i(x)\right]dt
\notag\\&\quad
 - a(x) r^{2-\theta} dr + \mathcal A_\mu(x) dx^\mu
 + \left(F_2(r) \partial_i r_0 + F_4(r) \partial_i b \right) (dx^i - v^i dt) \ ,
\end{align}
and $\phi_1$ has no correction term,
while $\phi_2$ receives the correction term $\varphi_2$;
\begin{align}
 \phi_2 &= - \frac{\theta}{(d-1)\tilde\nu} \log r
 + \frac{2}{\tilde\nu} \chi_0 + \varphi_2 \ .
\end{align}
The solution above must satisfy the following constraints;
\begin{align}
 0 &= \partial_t a + v^i \partial_i a - a \partial_i v^i  ,
\label{ConstADimRed}
\\
 0 &= \partial_t r_0 + v^i \partial_i r_0 + \frac{1}{3-\theta} r_0 \partial_i v^i  ,
\label{ConstTDimRed}
\\
 0 &= \mathcal F_{ti} + v^j \mathcal F_{ji}
 + \frac{z+3-\theta}{2 (z-1)} a b^{\frac{\theta}{3(3-\theta)}} r_0^{z+3-\theta}
 \left(z \frac{\partial_i r_0}{r_0} + \frac{\theta}{3(3-\theta)} \frac{\partial_i b}{b} \right)
\label{ConstSDimRed}
\end{align}
The hydrodynamic solution from the dimensional reduction also
reproduce the result in the previous sections, if $b = a$.

\subsection{Higher-dimensional thermodynamics and hydrodynamics}

We first consider thermodynamics before the dimensional reduction.
For $D$-dimensional space-time with arbitrary $D = d-\theta$,
the energy, entropy and charge are given by
\begin{align}
 E_D
 &=
 \mathcal E_D V_{D-1} = \frac{D-1}{16\pi G} r_0^{z+D-1} V_{D-1}
\\
 S_D
 &=
 s V_{D-1} = \frac{1}{4G} r_0^{D-1} V_{D-1} \ ,
\\
 N_D &= n V_{D-1} = \frac{z-1}{16 \pi G a} V_{D-1} \ ,
\end{align}
where $V_{D-1}$ is the volume of the $(D-1)$-dimensional space.
The 1st law of thermodynamics is expressed as
\begin{equation}
 dE_D = T_D dS_D - P_D dV_{D-1} + \mu_D dN_D \ ,
\end{equation}
and the temperature, pressure and chemical potential are calculated as
\begin{align}
 T_D &= \left(\frac{\partial E_D}{\partial S_D}\right)_{V_{D-1}, N_D}
 = \frac{D-1}{4\pi} r_0^z \ ,
\\
 P_D &= - \left(\frac{\partial E_D}{\partial V_{D-1}}\right)_{S_D, N_D}
 = \frac{z}{16\pi G} r_0^{z+D-1} \ ,
\\
 \mu_D &= \left(\frac{\partial E_D}{\partial N_D}\right)_{S_D, V_{D-1}}
 = 0 \ .
\end{align}
Here, the temperature $T_D$ agrees with that for the local observer
\begin{equation}
 T_D = \frac{T}{\sqrt{g_{tt}}} \ ,
\end{equation}
where $T$ is the Hawking temperature of the black hole.

Next, we consider the fluid in $D$-dimensional space-time with $D -1=3-\theta$.
The vielbein behaves near the boundary $r\to\infty$ as
\begin{align}
 E^0_{(D)} &= r^z \tau_D \ , &
 E^i_{(D)} &= r \hat e^i_D \ ,
\end{align}
where
\begin{align}
 \tau_D &= e^{-\chi_0} dt \ ,
\\
 \hat e^i_D &= e^{-\chi_0} \left(dx^i - v^i dt\right) \ , &
 (i &= 1,2,3)
\\
 \hat e^i_D &= e^{-\frac{6}{\theta}\chi_0} dx^i \ . &
 (i &= 4,\cdots,3-\theta)
\end{align}
This implies that the Newton-Cartan data on the boundary is given by
\begin{align}
 \tau_D &= e^{-\chi_0} dt \ ,
\\
 \hat v_D^\mu &= e^{\chi_0} (1,v^i,0) \ ,
\\
 h_D^{\mu\nu} &= \mathrm{diag}
 (0,
 e^{2\chi_0} , \cdots , e^{2\chi_0} ,
 e^{\frac{6}{\theta}\chi_0} , \cdots , e^{\frac{6}{\theta}\chi_0} ) \ ,
\end{align}

The stress-energy tensor on the boundary is calculated from
the first order solution for the metric before the dimensional reduction \eqref{SolBeforeDR},
in a similar fashion to Section~\ref{sec:StressTensor} as
\begin{align}
 \widetilde T_D{}^0{}_0
 &=
 \frac{1}{8\pi G}
 \left(- \frac{3-\theta}{2} r_0^{z+3-\theta}
 - \frac{z-1}{a} b^{-\frac{\theta}{3(3-\theta)}} v^i \mathcal A_i \right)
 \ ,
\\
 \widetilde T_D{}^i{}_0
 &=
 \frac{1}{8\pi G}
 \biggl[-\frac{z+3-\theta}{2} r_0^{z+3-\theta} v^i
 + \frac{z-1}{a} b^{-\frac{\theta}{3(3-\theta)}} v^i \mathcal A_t
 + \frac{1}{2} r_0^{3-\theta} \sigma_D{}^i{}_{j} v^j
\notag\\&\quad\qquad\qquad
 + \frac{z(z+3-\theta)}{4(z-1)} b^{-\frac{\theta}{3(3-\theta)}} r_0^{2z-\theta} \left( \partial_i r_0
 - \frac{\theta}{3 (3-\theta) z} \frac{r_0}{b} \partial_i b \right)
 \biggr]
 \ ,
\\
 \widetilde T_D{}^0{}_i
 &=
 \frac{1}{8\pi G}\frac{z-1}{a} b^{-\frac{\theta}{3(3-\theta)}}
 \mathcal A_i
 \ ,
\\
 \widetilde T_D{}^i{}_j
 &=
 \frac{1}{8\pi G}
 \biggl\{\frac{z}{2} r_0^{z+3-\theta} \delta_{ij}
 - \frac{1}{2} r_0^{3-\theta} \sigma_D{}^i{}_{j}
 + \frac{z-1}{a} b^{-\frac{\theta}{3(3-\theta)}} \left[v^i \mathcal A_j
 - \delta_{ij} \left(\mathcal A_t + v^k \mathcal A_k\right)\right]
 \biggr\} \ ,
\end{align}
where the shear tensor is given by
\begin{equation}
 \sigma_D{}^i{}_j
 = b^{-\frac{\theta}{3(3-\theta)}}
 \left(\partial_i v^j + \partial_j v^i - \frac{2}{3-\theta} \delta_{ij} \partial_k v^k \right)
 + \frac{2\theta}{3(3-\theta)}
 b^{-\frac{\theta}{3(3-\theta)}-1} \left(\partial_t b + v^k \partial_k b\right) \delta_{ij} \ ,
\end{equation}
for $i,j=1,\cdots,d-1$.

This stress-energy tensor can be expressed in the context of the Newton-Cartan theory as
\begin{align}
 \widetilde T_D{}^\mu{}_\nu
 &=
 \mathcal E_D \hat v_D^\mu \tau_D{}_\nu + P_D \hat h_D{}^\mu{}_\nu
 - \kappa_D \tau_D{}_\nu h_D^{\mu\rho} \left(\partial_\rho - \mathcal G_\rho\right) T_D
\notag\\&\quad
 - \eta_D \sigma_D{}_{\rho\sigma} h_D^{\rho\mu} \hat h_D{}^\sigma{}_\nu
 + n \hat v_D^\mu \mathcal A_\nu - n \hat v_D^\rho \mathcal A_\rho \delta^\mu{}_\nu \ ,
\end{align}
where we have defined
\begin{equation}
 \mathcal G_\mu = \left(\partial_\nu \tau_D{}_\mu- \partial_\mu \tau_D{}_\nu \right) \hat v_D^\nu \ .
\end{equation}
The shear tensor can be expressed in terms of the Newton-Cartan geometry as
\begin{align}
 \sigma_D{}_{\mu\nu}
 &=
 \hat h_D{}_{\rho\nu} \widehat D_\mu \hat v_D^\rho
 + \hat h_D{}_{\rho\mu} \widehat D_\nu \hat v_D^\rho
 - \frac{2}{3-\theta} \hat h_D{}_{\mu\nu} \widehat D_\rho \hat v_D^\rho
\notag\\
 &= \pounds_{\hat v_D} \hat h_D{}_{\mu\nu}
 - \frac{2}{3-\theta} \hat h_D{}_{\mu\nu} \widehat D_\rho \hat v_D^\rho \ ,
\end{align}
where $\widehat D_\mu$ is the covariant derivative in the Newton-Cartan geometry
in the holographic frame $\bar v^\mu = \hat v^\mu_D$,
(whose Christoffel symbol is given by \eqref{NCC} with $\bar v^\mu = \hat v_D^\mu$,
see Appendix~\ref{app:NewtonCartan} for more details),
and $\pounds$ is the Lie derivative.

The energy density, pressure, particle number density and the temperature are given by
\begin{align}
 \mathcal E_D &= \frac{3-\theta}{16\pi G} r_0^{z+3-\theta} \ , &
 P_D &= \frac{z}{16\pi G} r_0^{z+3-\theta} \ , &
 n &= \frac{z-1}{16\pi G} a^{-1}\ , &
 T_D &= \frac{z+3-\theta}{4\pi} r_0^z \ .
\end{align}
The transport coefficients are read off as
\begin{align}
 \eta_D &= \frac{1}{16\pi G} r_0^{3-\theta} \ , &
 \zeta_D &= 0 \ , &
 \kappa_D &= \frac{1}{8(z-1)G} r_0^{z+1-\theta} \ .
\end{align}

The stress-energy tensor can also be expressed in terms of the energy current,
momentum density, stress tensor and particle number current as
\begin{align}
 \widetilde T_D{}^\mu{}_\nu
 =
 - \widetilde{\mathcal E}_D^\mu \tau_D{}_\nu + \hat v_D^\mu \widetilde{\mathcal P}_D{}_\nu
 + \widetilde{\mathcal T}_D{}^\mu{}_\nu
 + J_D^\mu \mathcal A_\nu - J_D^\rho \mathcal A_\rho \delta^\mu{}_\nu
 \ ,
\end{align}
where
\begin{align}
 \widetilde{\mathcal E}_D^\mu
 &=
 \mathcal E_D \hat v_D^\mu
 - \kappa_D h_D^{\mu\rho} \left(\partial_\rho - \mathcal G_\rho\right) T_D \ ,
\\
 \widetilde{\mathcal P}_\nu
 &=
 0 \ ,
\\
 \widetilde{\mathcal T}_D{}^\mu{}_\nu
 &=
 P_D \hat h_D{}^\mu{}_\nu
 - \eta_D \sigma_D{}_{\rho\sigma} h_D{}^{\rho\mu} \hat h_D{}^\sigma_\nu \ ,
\\
 J_D^\mu
 &=
 n \hat v_D^\mu \ .
\end{align}

The conservation equations in the Newton-Cartan theory is given by
\begin{align}
 (\widehat D_\mu - \mathcal G_\mu) \widetilde{\mathcal E}_D^\mu
 &=
 \hat v_D^\mu (F^\tau_{\mu\nu} \widetilde{\mathcal E}_D^\nu - \mathcal F_{\mu\nu} J_D^\nu)
 - (\widehat D_\mu \hat v_D{}^\nu) \widetilde{\mathcal T}_D{}^\mu{}_\nu \ ,
\\
 h_D^{\rho\mu} (\widehat D_\nu - \mathcal G_\nu) \widetilde{\mathcal T}_D{}^\nu{}_\mu
 &=
 h_D^{\rho\mu} \left[\hat v_D^\nu \widehat D_\mu \widetilde{\mathcal P}_D{}_\nu
 - \widehat D_\nu (\hat v_D^\nu \widetilde{\mathcal P}_D{}_\mu)
 + \mathcal F_{\mu\nu} J_D^\nu - F^\tau_{\mu\nu} \widetilde{\mathcal E}_D^\nu \right] \ ,
\\
 0 &= \left(\widehat D_\mu - \mathcal G_\mu\right) J_D^\mu \ ,
\end{align}
where
\begin{align}
 F^\tau_{\mu\nu} &= \partial_\mu \tau_D{}_\nu - \partial_\nu \tau_D{}_\mu \ .
\end{align}
Then, they can be expressed in terms of the fluid variables as
\begin{align}
 0 &= \hat v_D^\mu \partial_\mu \mathcal E_D
 + (\mathcal E_D + P_D) \widehat D_\mu \hat v_D^\mu
 - \frac{1}{2} \eta_D \sigma_D{}^\mu{}_\nu \sigma_D{}^\nu{}_\mu
 - (\widehat D_\mu - 2 \mathcal G_\mu)
 \left[\kappa_D h_D^{\mu\rho} \left(\partial_\rho - \mathcal G_\rho\right) T_D\right]
\ ,
\\
 0 &= h_D^{\rho\nu} \partial_\nu P_D - h_D^{\rho\nu} \mathcal G_\nu (\mathcal E_D + P_D)
 - h_D^{\rho\mu} \mathcal F_{\mu\nu} J_D^\nu
 - h_D^{\mu\rho} h_D^{\nu\sigma}(\widehat D_\sigma - \mathcal G_\sigma)
 \left(\eta_D \sigma_D{}_{\mu\nu}\right)
\ ,
\\
 0 &= \widehat D_\mu \left(n \hat v_D^\mu\right) \ .
\end{align}
It is straightforward to verify that these equations are equivalent to
the constraint equations in the bulk equations of motion.

\subsection{Thermodynamics after the dimensional reduction}\label{sec:ThermoDR}

In this subsection we take the lower dimensional boundary  theory to have space dimension $d-1$. For general $b$, we consider in the following the first law, where the thermodynamic variables are the entropy $S$, the $(d-1)$-dimensional volume\footnote{Measured in the $d$-dimensional metric in the Einstein frame.} $V$ and the charge $N$, which is related to the variable $a$.
We have already found that (in general $d$)
\begin{align}
 E
 &=
 \mathcal E V = \frac{d-1-\theta}{16\pi G} b^\frac{\theta}{(d-1)(d-1-\theta)} r_0^{z+d-1-\theta} V
\\
 S
 &=
 s V = \frac{1}{4G} r_0^{d-1-\theta} V \ ,
\\
 N &= nV = \frac{z-1}{16 \pi G a} V \ .
\end{align}
In general, $b$ is not related to the particle number density $n$
and hence the energy does not depend on the charge.
Then the first law can be written as,
\begin{equation}
 dE = TdS - PdV + \mu dN \ ,
\end{equation}
where
\begin{align}
 T
 &= \left(\frac{\partial E}{\partial S}\right)_{V,N}
 = \frac{z+d-1-\theta}{4\pi} b^\frac{\theta}{(d-1)(d-1-\theta)} r_0^{z} \ ,
\\
 P
 &= - \left(\frac{\partial E}{\partial V}\right)_{S,N}
 = \frac{z}{16\pi G} b^\frac{\theta}{(d-1)(d-1-\theta)} r_0^{z+d-1-\theta} \ \ ,
\\
 \mu
 &= \left(\frac{\partial E}{\partial N}\right)_{S,V}
 = 0 \ .
\end{align}
The black hole solution also has another variable $b$,
which is related to the scalar source as
\begin{equation}
 \tilde \phi_2 = \tilde\nu \log b \ .
\end{equation}
Then, by taking into account this scalar source as an additional thermodynamic variable,
the first law of thermodynamics may be expressed as
\begin{equation}
 dE = TdS - PdV + \mu dN + \langle\widetilde{\mathcal O}_2\rangle d\tilde\phi_2 \ ,
\end{equation}
where $\langle\widetilde{\mathcal O}_2\rangle$ is the vev of the dual operator to the scalar $\phi_2$,
which is expressed as
\begin{equation}
 \langle\widetilde{\mathcal O}_2\rangle
 =
 \frac{\tilde\nu}{2} b^\frac{\theta}{(d-1)(d-1-\theta)} \left(\sum_{\mu=0}^{d-1} T_D{}^\mu{}_\mu
 +\frac{d-1}{\theta}\sum_{i=d}^{d-1-\theta} T_D{}^i{}_i\right)
 =
 - \frac{\tilde\nu}{2}\mathcal E - \tilde\nu \eta \bar\vartheta \ .
\end{equation}
We will discuss  this operator later.

In the previous section, we obtained the thermodynamics in the lower dimension, with one less variable.
To recover it from the one here we must take a codimension one section of the thermodynamic variables.
As argued already, the correct section involves taking  $b=a$ above.
With this constraint, we obtain

\begin{align}
 E
 &=
 \mathcal E V = \frac{d-1-\theta}{16\pi G} a^\frac{\theta}{(d-1)(d-1-\theta)} r_0^{z+d-1-\theta} V
\\
 S
 &=
 s V = \frac{1}{4G} r_0^{d-1-\theta} V \ ,
\\
 N &= nV = \frac{z-1}{16 \pi G a} V \ ,
\end{align}
and the first law takes the form
\begin{align}
 d E
 =
 T dS - \widetilde P dV
 + \tilde\mu dN \ .
\end{align}
The temperature, pressure and chemical potential are now given by
\begin{align}
 T
 &=
 \left(\frac{\partial E}{\partial S}\right)_{V,N}
 = \frac{z+d-1-\theta}{4\pi} a^\frac{\theta}{(d-1)(d-1-\theta)} r_0^{z} \ ,
\\
 \widetilde P &=
 - \left(\frac{\partial E}{\partial V}\right)_{S,N}
 = \frac{1}{16\pi G} \left(z-\frac{\theta}{d-1}\right)
 a^\frac{\theta}{(d-1)(d-1-\theta)} r_0^{z+d-1-\theta}  \ , \label{Ptilde}
\\
 \tilde\mu &= \left(\frac{\partial E}{\partial N}\right)_{S,V}
 = - \frac{\theta}{(d-1)(z-1)}
 a^{1 + \frac{\theta}{(d-1)(d-1-\theta)}} r_0^{z+d-1-\theta} \ . \label{muTilde}
\end{align}

Note that these are exactly the thermodynamic quantities and the first law  we obtained in the previous section (for $d=4$).

The Ward identity for the scaling symmetry is simply expressed as
\begin{equation}
 \left(z-\frac{\theta}{d-1}\right) \mathcal E = (d-1-\theta) P \ .
\end{equation}

\subsection{Hydrodynamics after the dimensional reduction}

We now set back $d=4$.
The stress-energy tensor and the fluid equations in lower dimensional theory can be calculated straightforwardly.
The stress-energy tensor is obtained as
\begin{align}
 \widetilde T^0{}_0
 &=
 \frac{1}{8\pi G}
 \left(- \frac{3-\theta}{2} b^{\frac{\theta}{3(3-\theta)}}  r_0^{z+3-\theta}
 - \frac{z-1}{a} v^i \mathcal A_i \right)
 \ ,
\\
 \widetilde T^i{}_0
 &=
 \frac{1}{8\pi G}
 \biggl[-\frac{z+3-\theta}{2} b^{\frac{\theta}{3(3-\theta)}} r_0^{z+3-\theta} v^i
 + \frac{z-1}{a} v^i \mathcal A_t
 + \frac{1}{2} r_0^{3-\theta} \sigma_{ij} v^j
\notag\\&\quad\qquad\qquad
 - \frac{\theta}{3(3-\theta)} r_0^{3-\theta}
 \left[\partial_j v^j - b^{-1}\left(\partial_t b + v^j \partial_j b\right)\right] v^i
\notag\\&\quad\qquad\qquad
 + \frac{z(z+3-\theta)}{4(z-1)} r_0^{2z-\theta} \left( \partial_i r_0
 - \frac{\theta}{3 (3-\theta) z} \frac{r_0}{b} \partial_i b \right)
 \biggr]
 \ ,
\\
 \widetilde T^0{}_i
 &=
 \frac{1}{8\pi G}\frac{z-1}{a}
 \mathcal A_i
 \ ,
\\
 \widetilde T^i{}_j
 &=
 \frac{1}{8\pi G}
 \biggl\{\frac{z}{2} b^{\frac{\theta}{3(3-\theta)}} r_0^{z+3-\theta} \delta_{ij}
 - \frac{1}{2} r_0^{3-\theta} \sigma_{ij}
 + \frac{z-1}{a} v^i \mathcal A_j - \delta_{ij} \left(\mathcal A_t + v^k \mathcal A_k\right)
\notag\\&\quad\qquad\qquad
 + \frac{\theta}{3(3-\theta)} r_0^{3-\theta}
 \left[\partial_i v^i - b^{-1}\left(\partial_t b + v^i \partial_i b\right)\right]
 \biggr\} \ ,
\end{align}
and the expectation values of the dual operators of the dilatons $\phi_1$ and $\phi_2$
are calculated as
\begin{align}
 \langle{\mathcal O}_1\rangle
 &= - \frac{\sqrt{(z-1)(3-\theta)}}{16 \pi G}
 \left[ \frac{1}{2} b^\frac{\theta}{3(3-\theta)} r_0^{z+3-\theta}
 - \frac{2\sqrt{2}}{a} \left( \mathcal A_t + v^i \mathcal A_i \right)
 \right]
 \ .
\\
 \langle{\mathcal O}_2\rangle
 &= - \frac{1}{16 \pi G}\biggl\{
 \sqrt{\frac{-\theta}{6(3-\theta)}}\left[ (z+2-\theta) b^\frac{\theta}{3(3-\theta)} r_0^{z+3-\theta}
 - \frac{2(z-1)}{a} \left( \mathcal A_t + v^i \mathcal A_i \right)
 \right]
\notag\\&\quad\qquad
 + \sqrt{-\frac{2\theta}{3(3-\theta)}}\,
 r_0^{3-\theta}\left[\partial_i v^i - b^{-1} (\partial_t b + v^i \partial_i b) \right]
 \biggr\}
\ .
\end{align}
For $b=a$, the results above agree with those in Section~\ref{sec:StressTensor}.

The stress-energy tensor after the dimensional reduction $\widetilde T^\mu{}_\nu$
can be expressed in terms of the fluid variables as
\begin{align}
 \widetilde T^\mu{}_\nu
 &=
 \mathcal E \hat v^\mu \tau_\nu + P \hat h^\mu{}_\nu
 - \kappa \tau_\nu h^{\mu\rho} \partial_\rho T
 - \eta \sigma_{ab} e_a^\mu \hat e^b_\nu
\notag\\&\quad
 - \zeta \bar\vartheta ~e_a^\mu \hat e^a_\nu
 + n \hat v^\mu \mathcal A_\nu - n \hat v^\rho \mathcal A_\rho \delta^\mu{}_\nu \ ,
 \label{StressEnergyDimRed}
\end{align}
where $\bar\vartheta$ is defined by
\begin{equation}
 \bar\vartheta = \partial_i v^i - b^{-1}\left(\partial_t b + v^i \partial_i b\right) \ .
\end{equation}
If $b=1$, this term gives the expansion term $\partial_i v^i$.
The energy density $\mathcal E$, pressure $P$, particle number density $n$
are the same as those in Section~\ref{sec:ThermoDR};
\begin{align}
 \mathcal E &= \frac{3-\theta}{16\pi G} b^{\frac{\theta}{3(3-\theta)}} r_0^{z+3-\theta} \ , &
 P &= \frac{z}{16\pi G} b^{\frac{\theta}{3(3-\theta)}} r_0^{z+3-\theta} \ , &
 n &= \frac{z-1}{16\pi G} a^{-1}\ .
\label{FluidVariablesDimRed}
\end{align}
The transport coefficients, heat conductivity $\kappa$ and shear viscosity $\eta$ are also read off as
\begin{align}
 \kappa &= \frac{1}{8(z-1)G} b^{-\frac{\theta}{3(3-\theta)}} r_0^{z+1-\theta} \ ,
&
 \eta &= \frac{1}{16\pi G} r_0^{3-\theta} \ .
\end{align}

For $b=a$, $\bar\vartheta$ vanishes but it does not mean that the fluid is incompressible,
since $\bar\vartheta$ is not only the expansion but has extra terms.
The expansion is non-zero but cancels with the extra terms, and hence,
this implies that the bulk viscosity vanishes in the lower dimension.
For the dimensional reduction with $b=1$ as in \eqref{MetricHigherDim},
$\bar\vartheta$  simply gives the expansion and hence the bulk viscosity $\zeta$ in that case is
\begin{equation}
 \zeta = - \frac{1}{8\pi G}\frac{\theta}{3(3-\theta)} r_0^{3-\theta} \ .
\end{equation}
It should be noted that $\theta$ is negative
for the dimensional reduction and hence $\zeta$ is positive.

The constraint equations can be written in terms of the fluid variables as
\begin{align}
 0 &= \partial_t \mathcal E + v^i \partial_i \mathcal E + (\mathcal E + P) \partial_i v^i
\notag\\&\qquad
 - \frac{1}{2} \eta \sigma_{ij}\sigma_{ij} - \zeta \bar\vartheta \partial_i v^i
 - \partial_i( \kappa \partial_i T)
 - \tilde\nu\,\langle\widetilde{\mathcal O}_2\rangle
 b^{-1} \left(\partial_t b + v^i \partial_i b \right)
\ , \label{LifContinuityDimRed}
\\
 0 &= \partial_i P + J^\mu \mathcal F_{\mu i}
  - \partial_j \left(\eta\sigma_{ij} \right)
  - \partial_i \left(\zeta\bar\vartheta\right)
  + \tilde\nu \,\langle\widetilde{\mathcal O}_2\rangle b^{-1} \partial_i b
\ , \label{LifNavierStokesDimRed}
\\
 0 &= \partial_t n + \partial_j (n v^j) \ , \label{LifChargeConsDimRed}
\end{align}
where $\widetilde{\mathcal O}_2$ is the dual of the dilaton $\phi_2$ but without the contribution from
the counter term $A^2$;
\begin{equation}
 \langle\widetilde{\mathcal O}_2\rangle = - \frac{1}{16 \pi G}
 \sqrt{\frac{-2\theta}{3(3-\theta)}}
 \left\{ \frac{3-\theta}{2} b^\frac{\theta}{3(3-\theta)} r_0^{z+3-\theta}
 + r_0^{3-\theta}\left[\partial_i v^i - b^{-1} (\partial_t b + v^i \partial_i b) \right] \right\}\ .
\end{equation}
This expression is consistent to the dimensional reduction from
the higher dimensional fluid;
\begin{align}
 \langle\widetilde{\mathcal O}_2\rangle
 &=
 \frac{\tilde\nu}{2} b^\frac{\theta}{3(3-\theta)} \left(\sum_{\mu=0}^3 T_D{}^\mu{}_\mu
 +\frac{3}{\theta}\sum_{i=4}^{3-\theta} T_D{}^i{}_i\right)
 =
 - \frac{\tilde\nu}{2}\mathcal E - \tilde\nu \eta \bar\vartheta \ ,
\end{align}
where $T_D{}^\mu{}_\nu$ is the stress-energy tensor of the fluid before the dimensional reduction.
Thus, $\langle\widetilde{\mathcal O}_2\rangle$ is related to
the other fluid variables if the fluid is obtained by the dimensional reduction.
It should be noted that the constant mode of $\phi_2$ is given by $\tilde\nu\log b$,
and hence,the contribution from $\widetilde{\mathcal O}_2$ is interpreted
as the coupling to the external source of $\phi_2$.
Contrary to the single dilaton case in the previous sections,
the variable $b$ is independent from the fluid variables and
is interpreted as an external field.
The transport coefficients also depend on temperature and external field $b$,
but are independent of the particle number density.

The energy $\mathcal E$, stress tensor $\widehat{\mathcal T}^i{}_j$ and
scalar operator $\widetilde{\mathcal O}_2$ satisfy
the following condition;
\begin{equation}
 0 =
 - \left(z - \frac{\theta}{d-1}\right) \mathcal E
 + \left(1 - \frac{\theta}{d-1}\right) \widehat{\mathcal T}^i{}_i
 - (d-1-\theta) \tilde\nu \langle\widetilde{\mathcal O}_2\rangle
\end{equation}
where the trace of the Milne invariant stress tensor is given by
\begin{equation}
 \widehat{\mathcal T}^i{}_i = (d-1) \left(P - \zeta \bar\vartheta\right) \ .
\end{equation}
The above condition is nothing but the Ward identity of
the Lifshitz scaling symmetry with the hyperscaling-violation.
The coefficients of $\mathcal E$, $P$ and $\tilde\nu\langle\widetilde{\mathcal O}_2\rangle$
equal to the scaling dimensions of $t$, $x^i$ and $b$ with appropriate signs, respectively.

If the fluid satisfies the condition $\bar\vartheta=0$,
the fluid equations \eqref{LifContinuityDimRed} and \eqref{LifNavierStokesDimRed}
can be rewritten as
\begin{align}
 0 &= \partial_t \mathcal E + v^i \partial_i \mathcal E + (\mathcal E + \widetilde P) \partial_i v^i
 - \frac{1}{2} \eta \sigma_{ij}\sigma_{ij}  - \partial_i( \kappa \partial_i T)
\ ,
\\
 0 &= \partial_i \widetilde P - \tilde n \partial_i \tilde\mu + J^\mu \mathcal F_{\mu i}
  - \partial_j \left(\eta\sigma_{ij} \right)
\ ,
\end{align}
where
\begin{align}
 \widetilde P &= P + \tilde \mu \tilde n
 = \frac{1}{16\pi G} \left(z-\frac{\theta}{d-1}\right)
 b^\frac{\theta}{(d-1)(d-1-\theta)} r_0^{z+d-1-\theta} \ ,
\label{EfPressure}
\\
 \tilde n &= \frac{(z-1)}{16\pi G b} \ ,
\\
 \tilde \mu &= - \tilde n^{-1} \tilde\nu\langle\widetilde{\mathcal O}_2\rangle
 = - \frac{\theta}{(d-1)(z-1)}
 b^{1 + \frac{\theta}{(d-1)(d-1-\theta)}} r_0^{z+d-1-\theta}
\ .
\label{2ndChemical}
\end{align}
The Ward identity of the Lifshitz scaling symmetry can also be expressed
in terms of $\widetilde P$ as
\begin{equation}
 \left(z-\frac{\theta}{d-1}\right) \mathcal E = (d-1-\theta) \widetilde P \ .
\end{equation}
For $b=a$, these equations agree with those in the previous section.
The above effective pressure \eqref{EfPressure} and chemical potential \eqref{2ndChemical}
agree with those in the thermodynamic relation \eqref{Ptilde} and \eqref{muTilde}, respectively.
Then, the thermodynamic relations, fluid equations and Ward identity,
as well as the stress-energy tensor reproduce the result in the previous section.

The result in this section is a generalization
of \cite{Kanitscheider:2009as} to non-relativistic and $z\neq 1$ cases.
The contributions from the gauge field and $\phi_1$
vanish for $z=1$ limit in which the Lifshitz black hole geometry becomes the Schwarzschild-AdS.
The hydrodynamic ansatz should be given by using the Lorentz boost for $z=1$,
and hence the fluid will be relativistic.
The results in this section are not well defined in $z=1$ limit,
since the ansatz is obtained by using the the Galilean boost.
The non-relativistic fluid which is obtained in this section
agrees with the non-relativistic limit of \cite{Kanitscheider:2009as} for $z=1$.

\section{Lifshitz hydrodynamics on a conformally flat background}\label{sec:NaiveAnsatz}

We have introduced the redefinition of the boundary coordinate before
replacing the parameters by slowly varying functions.
This coordinate redefinition is introduced to make
a flat space background on the boundary.
Here, we show that the naive hydrodynamic ansatz without
such a coordinate redefinition gives
fluids on a non-trivial but conformally flat background.

It can be calculated straightforwardly in a similar fashion to
Section~\ref{sec:HydroAnsatz} but without introducing
the coordinate redefinition \eqref{CoordRedef}.
Then, the first order constraint equations, which are
equivalent to the fluid equations in the perfect fluid limit,
are obtained as
\begin{align}
 0 &= \partial_t a + v^i \partial_i a - a \left(1-\frac{\theta}{3}\right) \partial_i v^i  ,
\label{ConstAConf}
\\
 0 &= \partial_t r_0 + v^i \partial_i r_0 + \frac13 r_0 \partial_i v^i  ,
\label{ConstTConf}
\\
 0 &= \mathcal F_{ti} + v^j\mathcal F_{ji} + \frac{z (z+3-\theta)}{2 (z-1)} a r_0^{z+2-\theta}  \partial_i r_0 \ .
\label{ConstSConf}
\end{align}
These equations are different from \eqref{ConstA}-\eqref{ConstS},
and as we will explain below, the difference can be interpreted as
the effect of the non-trivial background geometry at the boundary.
It is natural to expect that fluid variables as energy density, pressure and particle number density
are not affected by the background geometry.
In fact, we can calculate the stress-energy tensor straightforwardly and
they are read off as
\begin{align}
 \mathcal E &= \frac{3-\theta}{16\pi G} a^{\frac{\theta}{3(3-\theta)}} r_0^{z+3-\theta} \ , &
 P &= \frac{1}{16\pi G} \left(z-\frac{\theta}{d-1}\right)
 a^{\frac{\theta}{3(3-\theta)}} r_0^{z+3-\theta} \ ,
\notag\\
 n &= \frac{z-1}{8\pi G} a^{-1}\ , &
 \mu &= - \frac{\theta}{3(z-1)} a^{1 + \frac{\theta}{3(3-\theta)}} r_0^{z+3-\theta} \ .
\end{align}
which are the same as in \eqref{FluidVariablesZ}.

In order to consider the fluid mechanics in the non-trivial background,
we first introduce the fluid velocity field $u^\mu$ which is normalized as
\begin{equation}
 1 = \tau_\mu u^\mu
\end{equation}
where the timelike vielbein is given by
\begin{align}
 \tau &= e^{\chi_0} dt \ ,
\\
 e^{\chi_0} &= a^{-\frac{\theta}{3(3-\theta)}}
\end{align}
and hence the normalized velocity field $u^\mu$ is
\begin{align}
 u^t &= e^{-\chi_0} \ , &
 u^i &= e^{-\chi_0} v^i \ .
\end{align}
The constraint equations \eqref{ConstAConf}-\eqref{ConstSConf} can be expressed as
\begin{align}
 0 &= u^\mu \partial_\mu \mathcal E + (\mathcal E + P) D_\mu u^\mu
\ , \label{PFContConf}
\\
 0 &= \partial_i P - n \partial_i \mu + (\mathcal E + P) \partial_i \chi_0 + J^\mu \mathcal F_{\mu i}
\ ,
\\
 0 &= D_\mu J^\mu \ , \label{PFChargeConf}
\end{align}
where the particle number current $J^\mu$ is defined by
\begin{align}
 J^\mu &= n u^\mu \ .
\end{align}
Eq.~\eqref{PFContConf}-\eqref{PFChargeConf} are nothing but
the fluid equations in the perfect fluid limit in Newton-Cartan theory,
and the generalization to the first order fluid is straightforward.

The first order stress-energy tensor is obtained in the following form;
\begin{align}
 \widehat T^\mu{}_\nu
 &=
 \mathcal E u^\mu \tau_\nu + \left(P-n\mu\right) \hat h^\mu{}_\nu
 - \tilde\kappa \tau_\nu h^{\mu\rho} \partial_\rho T
 - \tilde\eta \sigma_{ab} e_a^\mu \hat e^b_\nu
 + n u^\mu \mathcal A_\nu - n u^\rho \mathcal A_\rho \delta^\mu{}_\nu \ ,
\end{align}
where the transport coefficients are given by
\begin{align}
 \tilde\kappa &= \frac{1}{8(z-1)G} r_0^{z+1-\theta} \ ,
&
 \tilde\eta &= \frac{1}{16\pi G} a^{\frac{\theta}{3(3-\theta)}} r_0^{3-\theta} \ ,
&
 \zeta &= 0 \ .
\end{align}
The bulk viscosity is zero as for the flat background \eqref{Transport}.
The difference of heat conductivity would come from the difference of the temperature.
In this case, the Hawking temperature is simply given by \eqref{HawkingT}
on the contrary to that in Section~\ref{sec:HydroAnsatz} where
the temperature is rescaled due to the coordinate transformation.
In the curved background, this should be expressed in terms of
the local temperature $T_T = e^{-\chi_0} T$ as
\begin{equation}
 \tilde\kappa\partial_\mu T = \kappa \left(\partial_\mu - \mathcal G_\mu\right) T_T \ ,
\end{equation}
and then, the heat conductivity is same as \eqref{Transport};
\begin{equation}
 \kappa = \frac{1}{8(z-1)G} a^{-\frac{\theta}{3(3-\theta)}} r_0^{z+1-\theta} \ ,
\end{equation}

The difference of the shear viscosity implies that
the shear tensor must be written in terms of
the normalized velocity field $u^\mu$;
\begin{equation}
 \hat\sigma_{\mu\nu}
 = \hat h_{\rho\nu} D_\mu u^\rho + \hat h_{\rho\mu} D_\nu u^\rho
 - \frac{2}{3} \hat h_{\mu\nu} D_\rho u^\rho \ .
\end{equation}
Then, the shear can be written as
\begin{equation}
 \tilde\eta \sigma_{ab} e_a^\mu \hat e^b_\nu = \eta \hat\sigma_{\rho\sigma} h^{\mu\rho} h^{\sigma}{}_\nu \ .
\end{equation}
Then, the shear viscosity $\eta$ equals to \eqref{Transport};
\begin{equation}
 \eta = \frac{1}{16\pi G} r_0^{3-\theta} \ .
\end{equation}

\subsection{Dimensional reduction for conformally flat background}

We can also consider the naive hydrodynamic ansatz for
the dimensional reduction from the higher dimensional Lifshitz geometry.
In this case, we can see the effects of the conformal factor in the metric more explicitly.
The perfect fluid limit of the fluid equations are obtained as
\begin{align}
 0 &= a^{-1}\left(\partial_t a + v^i \partial_i a \right)
 + \frac{\theta}{3-\theta} b^{-1}\left(\partial_t b + v^i \partial_i b\right) - \partial_i v^i  ,
\\
 0 &= r_0^{-1}\left(\partial_t r_0 + v^i \partial_i r_0\right)
 + \frac{\theta}{(3-\theta)^2} b^{-1}\left(\partial_t b + v^i \partial_i b\right)
 + \frac{1}{3-\theta} \partial_i v^i  ,
\\
 0 &= \mathcal F_{ti} + v^j \mathcal F_{ji}
 + \frac{z(z+3-\theta)}{2 (z-1)} a r_0^{z+2-\theta} \partial_i r_0 \ .
\end{align}
where $b$ comes from the effects of the non-trivial background;
\begin{equation}
 e^{\chi_0} \equiv b^{-\frac{\theta}{3(3-\theta)}} \ .
\end{equation}
The fluid variables are the same as in \eqref{FluidVariablesDimRed}
\begin{align}
 \mathcal E &= \frac{3-\theta}{16\pi G} b^{\frac{\theta}{3(3-\theta)}} r_0^{z+3-\theta} \ , &
 P &= \frac{z}{16\pi G} b^{\frac{\theta}{3(3-\theta)}} r_0^{z+3-\theta} \ , &
 n &= \frac{z-1}{8\pi G} a^{-1}\ ,
\end{align}
and the fluid equations can be written in terms of the normalized velocity field $u^\mu$
but now $\chi_0$ is independent from the particle number density $n\sim 1/a$;
\begin{align}
 0 &= u^\mu \partial_\mu \mathcal E + (\mathcal E + P) D_\mu u^\mu
 + \tilde\nu\widetilde{\mathcal O}_2 b^{-1} u^\mu \partial_\mu b
 \ ,
\\
 0 &= \partial_i P + (\mathcal E + P) \partial_i \chi_0 + J^\mu \mathcal F_{\mu i}
  - \tilde\nu\widetilde{\mathcal O}_2 b^{-1}\partial_i b
 \ ,
\\
 0 &= D_\mu J^\mu \ .
\end{align}
The first order stress-energy tensor is given by
\begin{align}
 \widehat T^\mu{}_\nu
 &=
 \mathcal E u^\mu \tau_\nu + P \hat h^\mu{}_\nu
 - \tilde\kappa \tau_\nu h^{\mu\rho} \partial_\rho T
 - \eta \hat\sigma_{\rho\sigma} h^{\mu\rho} h^{\sigma}{}_\nu
\notag\\&\quad
 - \zeta \left[\partial_i v^i - \frac{3}{3-\theta} b^{-1}\left(\partial_t b + v^i \partial_i b\right)\right] e_a^\mu \hat e^a_\nu
 + n u^\mu \mathcal A_\nu - n u^\rho \mathcal A_\rho \delta^\mu{}_\nu \ .
\end{align}
In this case, the expansion appears in the combination of
\begin{equation}
 \partial_i v^i - \frac{3}{3-\theta} b^{-1}\left(\partial_t b + v^i \partial_i b\right) \ ,
\end{equation}
and it vanishes for $b=a$ by substituting the constraint equation.
For $b=1$, the background becomes flat and hence
this agrees with \eqref{StressEnergyDimRed}.

\section*{Acknowledgements}
\addcontentsline{toc}{section}{Acknowledgements}

We would like to thank J. deBoer, N. Obers for discussions. We would like to thank especially  J. Hartong for enlightening discussions and for critical comments on the manuscript.

This work was supported in part by European Union's Seventh Framework Programme under grant agreements (FP7-REGPOT-2012-2013-1) no 316165 and the Advanced ERC grant SM-grav, No 669288.
The work is also supported in part by the Ministry of Science and Technology,
R.O.C. (project no. 104-2112-M-002 -003 -MY3) and by National Taiwan
University (project no. 105R8700-2).

\newpage
\appendix

\renewcommand{\theequation}{\thesection.\arabic{equation}}
\addcontentsline{toc}{section}{Appendices}
\section*{APPENDIX}

\section{Notations}\label{app:Notation}

\bigskip

\centerline{\bf Variables defined on the bulk (gravity) side}

\bigskip

\begin{itemize}

\item
$v^i$: Boost parameter introduced into the (static) black hole geometry in \eqref{BoostedBH}. It becomes the velocity of the fluid.

\item
$\mathcal A_\mu$: the constant part of the gauge field, which is defined in \eqref{gauge}.
This corresponds to the Milne boost-invariant gauge field
in the Newton-Cartan theory $\widehat B$, or equivalently,
$B$ in the holographic frame $\bar v^\mu = \hat v^\mu$.
In this paper, it is sometimes expressed as the 1-form $\mathcal A = \mathcal A_\mu dx^\mu$.

\item
$\tau_\mu$: the timelike vielbein on the boundary which is defined up to the factor of
$e^\chi r^{z}$, and given by \eqref{vl} in $r\to\infty$.
This corresponds to the timelike unit normal which defines the time direction in the Newton-Cartan theory.
It is automatically invariant under the Milne boost.

\item
$\hat e^a_\mu$: the spacelike vielbein on the boundary which is defined up to
the factor of $e^\chi r$, and given by \eqref{vl} in $r\to\infty$.
This corresponds to the spacelike vielbein in the Newton-Cartan theory
if we take the holographic frame $\bar v^\mu = \hat v^\mu$.

\item
$\hat v^\mu$: the timelike inverse vielbein on the boundary which is defined up to the factor
$e^{-\chi} r^{-z}$, and given by \eqref{hat} in $r\to\infty$.
This corresponds to the inverse timelike vielbein in the Newton-Cartan theory
if we take the holographic frame $\bar v^\mu = \hat v^\mu$, since
the holographic frame is defined by $\bar v^\mu = \hat v^\mu$.

\item
$e_a^\mu$: the spacelike inverse vielbein on the boundary which is defined up to the factor of
$e^{-\chi} r^{-1}$, and given by \eqref{hat} in $r\to\infty$.
This corresponds to the spacelike inverse vielbein in the Newton-Cartan theory.
It is automatically invariant under the Milne boost.

\item
$\widetilde T^\mu{}_\nu$:
the energy-momentum tensor calculated from the black hole geometry \eqref{SolGeom},
which is defined by \eqref{TtildeDef}

\item
$\widetilde{\mathcal E}^\mu$:
the energy current which is calculated from the black hole geometry \eqref{SolGeom}.
It is defined by \eqref{Ttilde2}.

\item
$\widetilde{\mathcal P}_\mu$:
the momentum density which is calculated from the black hole geometry \eqref{SolGeom}.
It is defined by \eqref{Ttilde2}.

\item
$\widetilde{\mathcal T}^\mu{}_\nu$:
the stress tensor which is calculated from the black hole geometry \eqref{SolGeom}.
It is defined by \eqref{Ttilde2}.

\item
$J^\mu$:
the particle number current which is calculated from the black hole geometry \eqref{SolGeom}.
It is given by \eqref{current}.

\end{itemize}

\bigskip

\centerline{\bf Variables defined in the (boundary) Newton-Cartan theory}

\bigskip

\noindent
The following variables in the Newton-Cartan theory are
appears in Section~\ref{sec:FluidEquation}.
See Appendix~\ref{app:NewtonCartan} for more details.

\begin{itemize}

\item
$\bar v^\mu$: the timelike inverse vielbein in Newton-Cartan theory.
The timelike inverse vielbein is not invariant under the Milne boost.
In the literature it is sometimes called  ``velocity'' but
must be distinguished from the velocity of the fluid.

\item
$h^{\mu\nu}$: the induced contravariant metric on the time-slice.
It is invariant under the Milne boost.

\item
$\bar h_{\mu\nu}$: the induced covariant metric on the time-slice.
It is not invariant under the Milne boost.

\item
$\hat h_{\mu\nu}$: the induced metric on the time-slice in the holographic frame
$\bar v^\mu = \hat v^\mu$. It is given by $\hat h_{\mu\nu} = \hat e^\mu_a \hat e^\nu_a$.
We also have $\hat h^\mu{}_\nu = h^{\mu\rho} \hat h_{\rho\nu}$ and
$\hat h^{\mu\nu} = h^{\mu\rho} \hat h_{\rho\sigma} h^{\sigma\nu} = h^{\mu\nu}$.

\item
$\bar e^a_\mu$: the spacelike vielbein.
It is given by $\bar e^a_\mu = \mathrm{diag}(0,1,1,1)$
for a 4-dim space-time. It satisfies $\bar h_{\mu\nu} = \bar e^a_\mu \bar e^a_\nu$.

\item
$\bar P^\mu{}_\nu$: the projection to the spatial direction,
which is defined by $e_a^\mu \bar e^a_\nu$, and satisfies
$\bar P^\mu{}_\nu = \bar h^\mu{}_\nu = h^{\mu\rho} \bar h_{\rho\nu}$.

\item
$u^\mu$: the fluid velocity field, $u^{\m}=(1,\vec v)$.
It equals to the vielbein $\hat v^\mu$, in the holographic frame .

\item
$B_\mu$: the gauge field in the Newton-Cartan theory.
It is not invariant under the Milne boost.

\item
$\mathcal E^\mu$: the energy current in the Newton-Cartan theory.
For fluids, it is given by \eqref{NewtonE}.

\item
$\mathcal P_\mu$: the momentum density in the Newton-Cartan theory.
For fluids, it is given by \eqref{NewtonP}.

\item
$\mathcal T^\mu{}_\nu$: the generic stress tensor in the Newton-Cartan theory.
For fluids, it is given by \eqref{NewtonT}.

\item
$\mathcal J^\mu$: the mass current in the Newton-Cartan theory.
For fluids, it is given by \eqref{Mcurrent}.

\item
$\bar T^\mu{}_\nu$: generic stress-energy tensor which is constructed from
$\mathcal E^\mu$, $\mathcal P_\mu$ and $\mathcal T^\mu{}_\nu$ in \eqref{Tbar}.

\item
$\widehat B$: a Milne boost invariant combination for the gauge field,
or equivalently, the gauge field in the holographic frame $\bar v^\mu = \hat v^\mu$.
It is defined by \eqref{HtoNewton}.

\item
$\widehat{\mathcal E}^\mu$: a Milne boost invariant combination for the energy current.
or equivalently, the energy current in the holographic frame $\bar v^\mu = \hat v^\mu$.
It is defined by \eqref{DefMinv}.

\item
$\widehat{\mathcal P}_\mu$: a Milne boost invariant combination for the stress tensor,
or equivalently, the momentum density in the holographic frame $\bar v^\mu = \hat v^\mu$.
It is defined by \eqref{DefMinv}.

\item
$\widehat{\mathcal T}^\mu{}_\nu$: a Milne boost invariant combination for the stress tensor,
or equivalently, the stress tensor in the holographic frame $\bar v^\mu = \hat v^\mu$.
It is defined by \eqref{DefMinv}.

\item
$T^\mu{}_\nu$: a Milne boost invariant combination of the stress-energy tensor,
or equivalently the stress-energy tensor in the holographic frame $\bar v^\mu = \hat v^\mu$.
It is defined by \eqref{TandTbar} and can be decomposed into
$\widehat{\mathcal E}^\mu$, $\widehat{\mathcal P}_\mu$ and $\widehat{\mathcal T}^\mu{}_\nu$.

\end{itemize}

\section{The Newton-Cartan formalism}\label{app:NewtonCartan}

Here, we briefly review the Newton-Cartan formalism.
The metric on the Galilei space-time is defined by
a 1-form $\tau_\mu$ and a contravariant symmetric tensor $h^{\mu\nu}$.
The time direction of the Galilei space-time is defined by the 1-form $\tau_\mu$.
$h^{\mu\nu}$ is the spatial inverse metric on the time-slice.
They satisfy the following orthogonality condition;
\begin{equation}
 \tau_\mu h^{\mu\nu} = 0 \ . \label{Orthogonality}
\end{equation}
Next, we introduce the covariant derivative $D_{\m}$ on the Galilei space-time.
We impose the condition that the Galilei data $(\tau_\mu,\ h^{\mu\nu})$
are constant under the covariant derivative
\begin{align}
 D_\mu \tau_\nu &= 0 \ ,
&
 D_\rho h^{\mu\nu} &= 0 \ .
\label{CovDer}
\end{align}
Unlike Einstein gravity,
the above conditions do not determine the Galilei connection uniquely.
We further introduce a contravariant vector $\bar v^\mu$,
which satisfies the following normalization condition;
\begin{equation}
 \tau_\mu \bar v^\mu = 1 \ . \label{Normalization}
\end{equation}
The vector $\bar v^\mu$ is the contravariant timelike vielbein. It is sometimes referred to as the velocity field
but is not to be confused with the velocity fluids.
We also define the spatial covariant metric $\bar h_{\mu\nu}$
by using the following conditions with $\bar v^\mu$;

\begin{align}
 \bar h_{\mu\nu} \bar v^\mu &= 0 \ ,
&
 \bar h_{\mu\rho} h^{\rho\nu} = \bar P^\nu{}_\mu = \delta^\nu{}_\mu- \bar v^\nu\tau_\mu \ .
\label{b4}\end{align}
We impose the spatial torsion free condition,
\begin{equation}
 \bar h_{\sigma\rho} T^\rho_{\mu\nu} = 0 \sp T^\rho_{\mu\nu} \equiv \Gamma^\rho_{\mu\nu} - \Gamma^\rho_{\nu\mu}\ .
\label{STfree}
\end{equation}
Then, the Newton-Cartan connection is determined up to a 2-form $H_{\mu\nu}$ as%
\begin{equation}
 \Gamma^\rho_{\mu\nu}
 =
 \bar v^\rho \partial_\mu \tau_\nu
 + \frac{1}{2} h^{\rho \sigma}
 \left(\partial_\mu\bar h_{\nu\sigma} + \partial_\nu\bar h_{\mu\sigma}
  - \partial_\sigma\bar h_{\mu\nu}\right)
 + \frac{1}{2} h^{\rho \sigma}
 \left(\tau_\mu H_{\nu \sigma} + \tau_\nu H_{\mu \sigma}\right)
 \ .
\label{NCC}
\end{equation}
In general, the Newton-Cartan connection has the torsion;
\begin{equation}
 T^\rho_{\mu\nu}
 = \bar v^{\rho} \left(\partial_\mu \tau_\nu - \partial_\nu \tau_\mu \right)
 \ .
\end{equation}
Without the spatial torsion free condition, (\ref{STfree}), the connection can be more general \cite{ho3}.

We can also define the Milne boost invariant connection \cite{Jensen:2014aia,ho3}
\begin{equation}
 \Gamma^\rho_{B\,\mu\nu}
 =
 v_B^\rho \partial_\mu \tau_\nu
 + \frac{1}{2} h^{\rho \sigma}
 \left(\partial_\mu h^B_{\nu\sigma} + \partial_\nu h^B_{\mu\sigma}
  - \partial_\sigma h^B_{\mu\nu}\right)
 \ .
\label{MNCC}
\end{equation}
where
\begin{align}
 v_B^\mu &= \bar v^\mu - h^{\mu\nu} B_\nu \ ,
&
 h^B_{\mu\nu} &= \bar h_{\mu\nu} + \tau_\mu B_\nu + \tau_\nu B_\mu \ ,
\end{align}
and $B$ is related to $H$ in \eqref{NCC} by $H = dB$, which we will discuss, soon.
The connection \eqref{MNCC} has non-zero spatial torsion;
\begin{equation}
 \bar h_{\sigma\rho} T^\rho_{B\,\mu\nu}
 = \bar h_{\sigma\rho} v_B^\rho \left(\partial_\mu \tau_\nu - \partial_\nu \tau_\mu\right)
 = - \bar P_\sigma^\rho B_\rho \left(\partial_\mu \tau_\nu - \partial_\nu \tau_\mu\right) \ ,
\end{equation}
or equivalently, \eqref{MNCC} satisfies the modified spatial torsion free condition,
$h^B_{\sigma\rho} T^\rho_{B\,\mu\nu} =0 $ instead of \eqref{STfree}.
Here, the spatial torsion free connection \eqref{NCC} is sufficient for our purposes
and hence we do not consider other connections as \eqref{MNCC}.

The curvature is defined by using the commutator of the covariant derivative and given by
\begin{equation}
 \mathcal R^\mu{}_{\nu\rho\sigma}
 =
 \partial_\rho \Gamma^\mu_{\nu\sigma} -  \partial_\sigma \Gamma^\mu_{\nu\rho}
 + \Gamma^\mu_{\alpha\rho} \Gamma^\alpha_{\nu\sigma}
 - \Gamma^\mu_{\alpha\sigma} \Gamma^\alpha_{\nu\rho}
 \ .
\end{equation}
For the torsion free case, $d\tau=0$,  the Newtonian condition can be imposed
\begin{equation}
 \mathcal R^{[\mu}{}_{(\nu}{}^{\rho]}{}_{\sigma)} = 0 \ ,
\label{NewtonCond}
\end{equation}
where $[\cdots]$ and $(\cdots)$ in the indices stand for
the antisymmetric part and symmetric part, respectively.
The above condition implies that
the 2-form $H_{\mu\nu}$ must be closed,
\be
dH=0\;.
\ee
Then, $H$ is interpreted as the field strength of a gauge field; $H = dB$.
For the torsional case, $d\tau\neq 0$, no appropriate generalization of the Newtonian condition is known.
Here, we simply impose the condition $H = dB$ even for $d\tau\neq 0$,
as was proposed in \cite{Jensen:2014aia}.

To summarize, the Newton-Cartan geometry is described by
the Newton-Cartan data, $\tau_\mu$, $h^{\mu\nu}$, $\bar v^\mu$ and $B_\mu$.

In the Newton-Cartan data, $\bar v^\mu$ and $B_\mu$
are introduced to define the Newton-Cartan connection.
Two pairs $(\bar v^\mu,\,B_\mu)$ and $(\bar v^{\prime\mu}, B_\mu')$ are equivalent
if they give the same Newton-Cartan connection.
The transformation from $(\bar v^\mu,\,B_\mu)$ to $(\bar v^{\prime\mu},\,B_\mu')$
gives an internal symmetry of the Newton-Cartan theory.
The Newton-Cartan connection is invariant under the following transformation, \cite{ho3};
\begin{align}
 \bar v^\mu &\to \bar v'{}^\mu = \bar v^\mu + h^{\mu\nu} V_\nu \ , \label{MilneV}\\
{B} &\to { B}' =
 { B} + \bar P_\mu^\nu V_\nu dx^\mu
 - \frac{1}{2} h^{\mu\nu} V_\mu V_\nu \tau_\rho dx^\rho \ .  \label{MilneA}
\end{align}
This transformation is known as the Milne boost and $V_\mu$ is a vector
which parametrizes the Milne boost.
It should be noted that the covariant spatial metric $\bar h_{\mu\nu}$ is
defined in terms of $\bar v^\mu$ and is not invariant under the Milne boost.
It transforms as
\begin{equation}
 \bar h'_{\mu\nu}
 =
 \bar h_{\mu\nu} - \left( \tau_\mu \bar P_\nu{}^\rho + \tau_\nu \bar P_\mu{}^\rho \right) V_\rho
 + \tau_\mu \tau_\nu h^{\rho \sigma} V_\rho V_\sigma
 \ . \label{MilneH}
\end{equation}

Next, we consider the conservation laws in the Newton-Cartan theory.
The energy current $\mathcal E^\mu$, momentum density $\mathcal P_\mu$,
stress tensor $\mathcal T_{\mu\nu}$ and mass current $\mathcal J^\mu$ are
given by the variation of the action with respect to
the Newton-Cartan data $\tau_\mu$, $h^{\mu\nu}$, $\bar v^\mu$ and $B_\mu$;
\begin{equation}
 \delta S
 =
 \int d^d x \sqrt{\gamma}
 \left[
  \delta \tau_\mu \mathcal E^\mu + \delta\bar v^\mu \mathcal P_\mu
  + \delta h^{\mu\nu} \mathcal T_{\mu\nu} + \delta B_\mu \mathcal J^\mu
 \right] \ ,
 \label{EPTJ}
\end{equation}
where the variation $\delta \tau_\mu$ is arbitrary but
$\delta\bar v^\mu$ and $\delta h^{\mu\nu}$ are
only the variations which satisfy the orthogonality and normalization conditions,
\eqref{Orthogonality} and \eqref{Normalization}, namely,
$\delta\bar v^\mu = \bar P^\mu_\nu \delta\bar v^\nu$, etc.
Invariance under coordinate transformations gives
the following conservation equations
\begin{align}
 \left(D_\mu - \mathcal G_\mu\right) \mathcal E^\mu
 &=
 \bar v^\mu \left(H_{\mu\nu} \mathcal J^\nu + K_{\mu\nu} \mathcal E^\nu\right)
 - \frac{1}{2}\left(h^{\mu\rho}D_\rho \bar v^\nu
 + h^{\nu\rho} D_\rho \bar v^\mu\right) \mathcal T_{\mu\nu} \ ,
\\
 h^{\rho\mu} h^{\sigma\mu}\left(D_\rho - \mathcal G_\rho\right) \mathcal T_{\mu\nu}
 &=
 h^{\sigma\nu}
 \left[
  \bar v^\mu D_\nu \mathcal P_\mu - D_\mu \left(\bar v^\mu \mathcal P_\nu\right)
  + H_{\mu\nu} \mathcal J^\mu + K_{\mu\nu} \mathcal E^\mu
 \right] \ ,
\end{align}
where
\begin{align}
 K_{\mu\nu} &= \partial_\mu \tau_\nu - \partial_\nu \tau_\mu \ ,
\\
 \mathcal G_\mu &= T^\nu_{\mu\nu} = - K_{\mu\nu} \bar v^\nu \ .
\end{align}
Invariance under the U(1) gauge symmetry of the gauge field $B_\mu$ gives
the conservation equation of the mass current;
\begin{equation}
 \left(D_\mu - \mathcal G_\mu\right) \mathcal J^\mu =0 \ .
\end{equation}
Invariance under the Milne boost gives the following Ward identity;
\begin{equation}
 \mathcal P_\mu = \bar h_{\mu\nu} \mathcal J^\nu  \ . \label{WardMilne}
\end{equation}

The energy flow, momentum density and stress tensor are not invariant under the Milne boost.
The variations transform under the Milne boost as
\begin{align}
 \delta\bar v^{\prime\mu}
 &=
 \delta\bar v^\mu + V_\nu \delta h^{\mu\nu} \ ,
\\
 \delta B'_\mu
 &=
 \delta B_\mu - \frac{1}{2} h^{\rho\nu} V_\rho V_\nu \delta \tau_\mu
 - \tau_\mu \left(V_\nu \delta\bar v^\nu + \frac{1}{2} V_\nu V_\rho \delta h^{\nu\rho}\right) \ ,
\end{align}
where we imposed $V_\mu \bar v^\mu=0$ since $V_\mu$ always appears together  with $h^{\mu\nu}$ or $\bar P^\mu{}_\nu$.
It should also be noted that the variations in \eqref{EPTJ} are constrained by
the orthogonality \eqref{Orthogonality} and normalization \eqref{Normalization} conditions
and hence give additional contributions proportional to the variation of $\tau_\mu$;
\begin{align}
 \delta \bar v^\mu
 &=
 - \bar v^\mu \bar v^\nu \delta \tau_\nu
 + \bar P^\mu_\nu \delta\bar v^\nu \ ,
\\
 \delta h^{\mu\nu}
 &=
 - \left( \bar v^\mu h^{\nu\rho} + \bar v^\nu h^{\mu\rho}\right) \delta \tau_\rho
 + \bar P^\mu_\rho \bar P^\nu_\sigma \delta h^{\rho\sigma} \ ,
\end{align}
where the first terms in these equations are already included in \eqref{EPTJ} but
their transformation under a Milne boost gives additional terms in the Milne transformation of the energy flow.
Then, the transformation of the energy flow, momentum density and stress tensor are given by
\begin{align}
 \mathcal E^{\prime\mu}
 &=
 \mathcal E^\mu - h^{\mu\rho} h^{\nu \sigma} \mathcal T_{\rho\nu} V_\sigma
 - \bar v^\mu h^{\nu\rho} \mathcal P_\nu V_\rho
 + \frac{1}{2} \mathcal J^\mu h^{\rho \sigma} V_\rho V_\sigma \ ,
 \label{MilneE}
\\
 \mathcal P'_\mu
 &=
 \mathcal P_\mu
 - \tau_\mu h^{\nu\rho} \mathcal P_\nu V_\rho
 - \tau_\rho \mathcal J^\rho \bar P_\mu^\nu V_\nu
 + \tau_\mu \tau_\nu \mathcal J^\nu h^{\rho \sigma} V_\rho V_\sigma
 \ , \label{MilneP}
\\
 \mathcal T^{\prime\mu\nu}
 &=
 \mathcal T^{\mu\nu} - \left(\mathcal P^\mu h^{\nu\rho} + \mathcal P^\nu h^{\mu\rho}\right) V_\rho
 + h^{\mu\rho} h^{\nu\sigma} V_\rho V_\sigma \tau_\lambda \mathcal J^\lambda \ , \label{MilneT}
\end{align}
where the indices are raised or lowered by the spatial metric $h^{\mu\nu}$ or $\bar h_{\mu\nu}$,
respectively, and the transformation of the stress tensor is given by \eqref{MilneT} with \eqref{MilneH}.
The mass current is invariant under the Milne boost;
\begin{equation}
 \mathcal J^{\prime\mu} = \mathcal J^\mu \ , \label{MilneJ}
\end{equation}
and in fact, \eqref{MilneP} is consistent with the Ward identity \eqref{WardMilne}
with \eqref{MilneH} and \eqref{MilneJ}.

We now introduce Milne boost invariant combinations.
It is straightforward to see that the following combination is invariant under the Milne boost;
\begin{equation}
 \widetilde{\mathcal T}^\mu{}_\nu
 = \mathcal J^\mu \bar v^\nu + \bar v^\mu \mathcal P^\nu + \mathcal T^{\mu\nu} \ ,
\end{equation}
which is an analogue of the stress-energy tensor, but constructed
from the mass current, momentum density and stress tensor.
We can also define the stress-energy tensor from
the energy flow, momentum density and stress tensor;
\begin{equation}
 \bar T^\mu{}_\nu
 = - \mathcal E^\mu \tau_\nu + \bar v^\mu \mathcal P_\nu
 + h^{\mu\rho} \mathcal T_{\rho\nu} \ ,
\end{equation}
but this stress-energy tensor is not invariant under the Milne boost
and transforms as
\begin{equation}
 \bar T^{\prime\mu}{}_\nu
 = \bar T^\mu{}_\nu - \mathcal J^\mu V_\nu
 + \frac{1}{2} \mathcal J^\mu \tau_\nu h^{\rho \sigma} V_\rho V_\sigma \ .
\end{equation}
With this stress-energy tensor, a Milne boost invariant combination is given by
\begin{equation}
 \widehat T^\mu{}_\nu
 =
 \bar T^\mu{}_\nu
 + \mathcal J^\mu B_\nu \ . \label{That}
\end{equation}
It should be noted that the Milne boost combination $\widehat T^\mu{}_\nu$
is not invariant (or covariant) under the U(1) gauge transformation associated to $B_\mu$.

Now, we consider fluids in the Newton-Cartan theory.
We introduce the fluid velocity field $u^\mu$, which is invariant under the Milne boost.
For the first order fluid in the Eckart frame,
the energy flow, momentum density, stress tensor and mass current are given by
\begin{align}
 \mathcal E^\mu
 &=
 \mathcal E u^\mu + \frac{1}{2} \rho u^2 u^\mu - \kappa h^{\mu\nu} \partial_\nu T
 + h^{\mu\rho} u^\sigma \mathcal T_{\rho \sigma} \ ,
\\
 \mathcal P_\mu &= \rho u_\mu \ ,
\\
 \mathcal T_{\mu\nu}
 &=
 P \bar h_{\mu\nu} + \rho u_\mu u_\nu
 - \eta \sigma_{\rho \sigma} \bar P_\mu^\rho \bar P_\nu^\sigma
 - \zeta \theta \bar h_{\mu\nu} \ ,
\\
 \mathcal J^\mu
 &=
 \rho u^\mu \ ,
\label{EckartJ}
\end{align}
where the index of the velocity field is lowered by using the spatial metric $\bar h_{\mu\nu}$
and $u^2$ is defined as
\begin{align}
 u_\mu &= \bar h_{\mu\nu} u^\nu \ ,
\\
 u^2 &= \bar h_{\mu\nu} u^\mu u^\nu \ .
\end{align}
The energy density $\mathcal E$, pressure $P$ and mass density $\rho$ are invariant under the Milne boost.
We construct a stress-energy tensor from the above quantities,
and then $\bar T^\mu{}_\nu$ is obtained as
\begin{align}
 \bar T^\mu{}_\nu
 &=
 - \mathcal E u^\mu \tau_\nu - \frac{1}{2} \rho u^2 u^\mu
 + P \widehat P^\mu_\nu
 + \rho u^\mu u_\nu
\notag\\&\quad
 - \eta \sigma_{\rho \sigma} h^{\mu\rho} \widehat P_\nu^\sigma
 - \zeta \theta \widehat P^\mu_\nu
 + \kappa h^{\mu\rho} \left(\partial_\rho T\right) \tau_\nu \ , \label{PhysT}
\end{align}
where
\begin{align}
 \widehat P^\mu{}_\nu \equiv \delta^\mu{}_\nu - u^\mu \tau_\nu
 = \bar P^\mu{}_\nu - h^{\mu\rho} u_\rho \tau_\nu \ .
\end{align}
We consider the Milne boost invariants in fluid mechanics.
We introduce a Milne boost invariant combination for the gauge field by using $u^\mu$;
\begin{align}
 \widehat B = B + u_\mu dx^\mu
 - \frac{1}{2} u^2 \tau_\rho dx^\rho \ . \label{Binv}
\end{align}
Then, the Milne boost invariant combination for the stress-energy tensor $\widehat T^\mu{}_\nu$
is expressed in terms of the Milne boost invariant gauge field $\widehat B_\mu$ as
\begin{align}
 \widehat T^\mu{}_\nu
 &=
 \bar T^\mu{}_\nu + \mathcal J^\mu B_\nu
\notag\\
 &=
 - \mathcal E u^\mu \tau_\nu
 + P \widehat P^\mu{}_\nu
 + \mathcal J^\mu \widehat B_\nu
 - \eta \sigma_{\rho \sigma} h^{\mu\rho} \widehat P^\sigma{}_\nu
 - \zeta \theta \widehat P^\mu{}_\nu
 + \kappa h^{\mu\rho} \left(\partial_\rho T\right) \tau_\nu \ . \label{T+JB}
\end{align}
This stress-energy tensor is invariant under the Milne boost,
but is not invariant under the U(1) gauge symmetry associated to $B_\mu$.
We define the Milne boost invariant and U(1) gauge invariant
stress-energy tensor, $T^\mu{}_\nu$ from $\widehat T^\mu{}_\nu$ and $\widehat B_\mu$ 
such that
\begin{equation}
 T^\mu{}_\nu \equiv \widehat T^\mu{}_\nu
 - \mathcal J^\mu \widehat B_\nu \ .
\end{equation}
Then, it is expressed as
\begin{align}
 T^\mu{}_\nu
 &=
 - \mathcal E u^\mu \tau_\nu
 + P \widehat P^\mu{}_\nu
 - \eta \sigma_{\rho \sigma} h^{\mu\rho} \widehat P^\sigma{}_\nu
 - \zeta \theta \widehat P^\mu{}_\nu
 + \kappa h^{\mu\rho} \left(\partial_\rho T\right) \tau_\nu \ . \label{MilneInvT}
\end{align}
The Milne boost invariant stress-energy tensor \eqref{MilneInvT}
can be decomposed into the Milne boost invariant energy density $\widehat{\mathcal E}^\mu$,
momentum density $\widehat{\mathcal P}_\mu$ and stress tensor $\widehat{\mathcal T}^\mu{}_\nu$ as
\begin{align}
 T^\mu{}_\nu
 &=
 - \widehat{\mathcal E}^\mu \tau_\nu + u^\mu \widehat{\mathcal P}_\nu
 + \widehat{\mathcal T}^\mu{}_\nu
\end{align}
where
\begin{align}
 \widehat{\mathcal E}^\mu
 &\equiv
 - T^\mu{}_\nu u^\nu
 =
 \mathcal E u^\mu - \kappa h^{\mu\nu} \partial_\nu T
\\
 \widehat{\mathcal P}_\mu
 &\equiv
 T^\rho{}_\nu \tau_\rho \widehat P^\nu{}_\mu
 = 0 \ ,
\\
 \widehat{\mathcal T}^\mu{}_\nu
 &\equiv
 T^\rho{}_\sigma \widehat P^\mu{}_\rho \widehat P^\sigma{}_\nu
 = P \widehat P^\mu{}_\nu
 - \eta \sigma_{\rho \sigma} h^{\mu\rho} \widehat P^\sigma{}_\nu
 - \zeta \theta \widehat P^\mu{}_\nu \ .
\end{align}
By using the relation between $T^\mu{}_\nu$ and $\bar T^\mu{}_\nu$;
\begin{equation}
 T^\mu{}_\nu = \bar T^\mu{}_\nu + \mathcal J^\mu B_\nu - \mathcal J^\mu \widehat B_\nu
 = \bar T^\mu{}_\nu - \mathcal J^\mu \left(u_\nu - \frac{1}{2}\tau_\nu u^2\right) ,
\end{equation}
the Milne boost invariant energy flow, momentum density and stress tensor are expressed as
\begin{align}
 \widehat{\mathcal E}^\mu
 &=
 \mathcal E^\mu - \mathcal T^\mu{}_\nu u^\nu - \bar v^\mu \mathcal P_\nu u^\nu
 + \frac{1}{2} u^2 \mathcal J^\mu \ ,
\label{MilneInvEnergy}
\\
 \widehat{\mathcal P}_\mu
 &=
 \left(\mathcal P_\nu - \tau_\rho \mathcal J^\rho u_\nu\right) \widehat P^\nu{}_\mu
 = 0
\label{MilneInvMomentum}
\\
 \widehat{\mathcal T}^\mu{}_\nu
 &=
 h^{\mu\rho} \left(\mathcal T_{\rho \sigma} - \rho u_\rho u_\sigma\right) \widehat P^\sigma{}_\nu
\label{MilneInvStress}
\end{align}
where we have used the condition for the Eckart frame \eqref{EckartJ} in the last equality of
\eqref{MilneInvMomentum} and in \eqref{MilneInvStress}.
The Milne boost invariant energy flow \eqref{MilneInvEnergy}
is equivalent to that introduced in \cite{Jensen:2014ama}

In general, the Milne boost invariant quantities $\widehat{\mathcal E}^\mu$,
$\widehat{\mathcal P}_\mu$, $\widehat{\mathcal T}^\mu{}_\nu$ and $\widehat T^\mu{}_\nu$
are different from the physical quantities ${\mathcal E}^\mu$,
${\mathcal P}_\mu$, ${\mathcal T}^\mu{}_\nu$ and $T^\mu{}_\nu$,
but they agree if we take a special frame of the Milne boost transformation
by imposing the condition
\begin{equation}
 \bar v^\mu = u^\mu \ ,
\end{equation}
and we can take this frame by using the Milne boost with
\begin{equation}
 V_\mu = \bar h_{\mu\nu} \left( u^\nu - \bar v^\nu \right) = u_\mu \ .
\end{equation}
This also implies that the conservation law for
the Milne boost invariant quantities $\widehat{\mathcal E}^\mu$,
$\widehat{\mathcal P}_\mu$ and $\widehat{\mathcal T}^\mu{}_\nu$
is given by
\begin{align}
 \left(\widehat D_\mu - \widehat{\mathcal G}_\mu\right) \widehat{\mathcal E}^\mu
 &=
 u^\mu \left(\widehat H_{\mu\nu} \mathcal J^\nu + K_{\mu\nu} \widehat{\mathcal E}^\nu\right)
 - \frac{1}{2}\left(h^{\mu\rho} \widehat D_\rho u^\nu
 + h^{\nu\rho} \widehat D_\rho u^\mu\right) \widehat{\mathcal T}_{\mu\nu} \ ,
\\
 h^{\rho\mu} h^{\sigma\nu}\left(\widehat D_\rho - \widehat{\mathcal G}_\rho\right)
 \widehat{\mathcal T}_{\mu\nu}
 &=
 h^{\sigma\nu}
 \left[
  u^\mu \widehat D_\nu \widehat{\mathcal P}_\mu - \widehat D_\mu \left( u^\mu \widehat{\mathcal P}_\nu\right)
  + \widehat H_{\mu\nu} \mathcal J^\mu + K_{\mu\nu} \widehat{\mathcal E}^\mu
 \right] \ ,
\end{align}
where $\widehat D_\mu$ is the covariant derivative in this frame and
\begin{equation}
 \widehat{\mathcal G}_\mu = - K_{\mu\nu} u^\mu \ .
\end{equation}
We call this frame the ``holographic frame,'' since
if we consider the holographic duality for the Lifshitz space-times, the hydrodynamic   results are obtained in this frame.

\section{Lifshitz fluid in scalar background}

In section~\ref{sec:FluidEquation}, we have seen that
the Lifshitz black hole geometry corresponds to
fluids with a non-zero chemical potential
but without non-trivial scalar background.
On the other hand, the dimensional reduction from the higher dimensional model in Section~\ref{sec:DimRed}
gives fluids with zero chemical potential but with non-trivial scalar background.
Here, we show that the results in Section~\ref{sec:FluidEquation} can also be
interpreted as fluids with zero chemical potential but with non-trivial scalar background,
even though the pressure is different from that in first law of thermodynamics.

The stress-energy tensor $\widetilde T^\mu{}_\nu$
can be expressed in the following form;
\begin{align}
 \widetilde T^\mu{}_\nu
 &=
 - \mathcal E \hat v^\mu \tau_\nu + P \hat h^\mu{}_\nu
 - \kappa \tau_\nu h^{\mu\rho} \partial_\rho T
 - \eta \sigma_{ab} e_a^\mu \hat e^b_\nu
 + n \hat v^\mu \mathcal A_\nu - n \hat v^\rho \mathcal A_\rho \delta^\mu{}_\nu \ ,
\end{align}
where
the energy density $\mathcal E$ and particle number density $n$
have the same form \eqref{FluidVariablesZ},
but the pressure is now given by
\begin{align}
 P &= \frac{z}{16\pi G} a^{\frac{\theta}{3(3-\theta)}} r_0^{z+3-\theta} \ .
\end{align}
The stress-energy tensor $\widetilde T^\mu{}_\nu$ is related to
that in the Newton-Cartan theory $T^\mu{}_\nu$ by
the same equation \eqref{TandTtilde},
but now the relation between $\mathcal A$ and $\widehat B$ is given by
\begin{equation}
 \mathcal A_\mu = m \widehat B_\mu \ ,
\end{equation}
which implies that $\mu = 0$.

The fluid equations are expressed as
\begin{align}
 0 &= \partial_t \mathcal E + v^i \partial_i \mathcal E + (\mathcal E + P) \partial_i v^i
\notag\\&\qquad
 - \frac{1}{2} \eta \sigma_{ij}\sigma_{ij}
 - \partial_i( \kappa \partial_i T)
 + c_\phi \,\langle \widetilde{\mathcal O}_\phi\rangle \frac{1}{n}\left(\partial_t n + v^i \partial_i n\right) \ ,
\label{ContNTS}
\\
 0 &= \partial_i P + n \mathcal F_{ti} + n v^j \mathcal F_{ji}
  - \partial_j \left(\eta\sigma_{ij} \right)
  - c_\phi \,\langle \widetilde{\mathcal O}_\phi\rangle  \frac{1}{n}\partial_i n
\ ,
\label{NSNTS}
\\
 0 &= \partial_t n + \partial_j (n v^j) \ ,
\end{align}
where $c_\phi$ is the coupling constant between $\widetilde{\mathcal O}_\phi$ and $\log a$,
which is given by
\begin{equation}
 c_\phi = - \frac{\theta}{3} \left\{
 \frac{9(z-1)+(3-\theta)\theta}{\sqrt{6(3-\theta)[3(z-1)-\theta]}}\right\}^{-1} \ , \label{cphi}
\end{equation}
and the operator $\widetilde{\mathcal O}_\phi$ is the dual of
the dilaton $\phi$ but we have removed the contribution form the external gauge field
as for the stress-energy tensor;
\begin{equation}
 \langle \widetilde{\mathcal O}_\phi\rangle
 = \langle \mathcal O_\phi\rangle - \left(\lambda-\frac{1}{2}\nu\right) J^\mu\mathcal A_\mu \ .
 \label{Otilde}
\end{equation}
where $\lambda$ and $\nu$ are defined in (\ref{la}) and (\ref{nu}) respectively.

The additional term associated to  $\widetilde{\mathcal O}_\phi$ can be understood as follows.
In terms of the boundary stress-energy tensor,
\eqref{ContNTS} and \eqref{NSNTS} are expressed as
\begin{equation}
 D_\mu {\cal T}^\mu{}_\nu = {\cal F}_{\mu\nu} J^\mu
 + \bar c_\phi \langle\widetilde{\mathcal O}_\phi\rangle \partial_\nu \phi_0 \ ,
\end{equation}
where $\phi_0$ is the constant mode of the dilaton $\phi$, which is defined in \eqref{Phi0} and
is interpreted as the source for $\widetilde O_\phi$.
The coupling constant $\bar c_\phi$ is now redefined as
\begin{align}
 \bar c_\phi &= -  \frac{(3-\theta)\theta}{9(z-1)+(3-\theta)\theta} \ .
\end{align}
This additional term implies the presence of an external force
whose potential is proportional to the particle number density.
Although, $\phi_0$ is interpreted as the external field,
it is related to the particle number density $n$,
or equivalently $a$, as seen from \eqref{Phi0}.

For the fluid dual to the Lifshitz geometry,
the energy density is constrained first by the Lifshitz Ward identity with hyperscaling violation. It is  given by
\begin{equation}
 - \left(z-\frac{\theta}{3}\right)\mathcal E + (3-\theta) P
 - (3-\theta) c_\phi \langle\widetilde{\mathcal O}_\phi\rangle = 0 \ ,
\end{equation}
where the operator $\widetilde{\mathcal O}_\phi$ and coupling constant $c_\phi$
are defined by \eqref{Otilde} and \eqref{cphi}, respectively.
The coefficients of $\mathcal E$, $P$ and $c_\phi\langle\widetilde{\mathcal O}_\phi\rangle$
are  the scaling dimensions of $t$, $x^i$ and $a$
with appropriate signs, respectively.
(See Appendix~\ref{app:ScaleDim} for more details
on the scaling dimensions for the Lifshitz symmetry with hyperscaling-violation.)

The thermodynamics relation given by
\begin{align}
 T J_S^\mu = \widehat{\mathcal{E}}^\mu + P \hat v^\mu
 = - T^\mu{}_\nu \hat v^\nu\ + P \hat v^\mu \ .
\end{align}
This is consistent with $\mu=0$.

\section{Scaling dimensions in fluids}\label{app:ScaleDim}

If there is no hyperscaling-violation,
the Lifshitz space-time is invariant under the following scaling transformation;
\begin{align}
 t &\to c^z t \ , &
 x^i &\to c x^i \ , &
 r &\to c^{-1} r \ .
\end{align}
However, in the presence of the hyperscaling-violation,
the following metric and gauge field are not invariant under the above transformation;
\begin{align}
 ds^2 &= b^{-\frac{2\theta}{(d-1)(d-1-\theta)}} r^{-2\theta/(d-1)}
 \left(- r^{2z} dt^2 + \frac{dr^2}{r^2} + \sum_i r^2 (dx^i)^2\right)
\\
 A &= a r^{z+d-1-\theta} dt \ ,
\end{align}
but transform as
\begin{align}
 ds^2 &\to c^{2\theta/(d-1)} ds^2 \ ,
&
 A &\to c^{-(d-1)+\theta} A \ .
\end{align}
where $a,b$ are constant parameters above.
However, if we impose the following scaling behavior to the constant parameters
\begin{align}
 a &\to c^{d-1-\theta} a \ , &
 b &\to c^{d-1-\theta} b \ ,
\end{align}
then the metric and gauge field are invariant under the scaling transformation.
For the black hole geometry, the horizon radius transforms as
\begin{equation}
 r_0 \to c^{-1} r_0 \ .
\end{equation}
Then, the coordinates and the constants have the following scaling dimensions:
\begin{align}
 [t] &= -z \ ,
&
 [x^i] &= -1 \ ,
\\
 [a] &= -(d-1) + \theta \ ,
&
 [b] &= -(d-1) + \theta \ ,
&
 [r_0] &= 1 \ .
\end{align}
After the redefinition of the coordinate \eqref{CoordRedef},
the scaling dimension of the coordinates becomes
\begin{align}
 [t] &= -z + \frac{\theta}{d-1}\ ,
&
 [x^i] &= -1 + \frac{\theta}{d-1} \ .
\end{align}
Then, the scaling dimension of the velocity field should be
\begin{equation}
 [v^i] = z-1 \ .
\end{equation}
The constant modes of the gauge field have the following dimensions
\begin{align}
 [\mathcal A_t] &= z-\frac{\theta}{d-1} \ , &
 [\mathcal A_i] &= 1-\frac{\theta}{d-1} \ .
\end{align}
{}From \eqref{FluidVariablesZ} and \eqref{HawkingTres}, or
\eqref{FluidVariablesDimRed} and \eqref{HawkingTdr},
the scaling dimensions of the energy density, pressure,
the particle number density and temperature are given by
\begin{align}
 [\mathcal E] &= z + (d-1) - \frac{d}{d-1} \theta \ , &
 [P] &= z + (d-1) - \frac{d}{d-1} \theta \ , &
\notag\\
 [n] &= d-1 - \theta \ , &
 [T] &= z - \frac{\theta}{d-1} \ .
\end{align}
They satisfies the appropriate relation to the dimension of the coordinates
\begin{align}
 [\mathcal E] &= - [t] - (d-1) [x^i] \ , &
 [n] &= - (d-1) [x^i] \ , &
 [T] &= - [t] \ ,
\end{align}
namely, dimensions of the density is minus of dimension of volume, and
dimensions of the energy and temperature are minus of the dimension of time.
{}From the relation to fluid variables,
the scaling dimensions of the transport coefficients are calculated as
\begin{align}
 [\eta] &= d-1-\theta \ , &
 [\zeta] &= d-1-\theta \ , &
 [\kappa] &= z-1-\frac{d-2}{d-1} \theta \ ,
\end{align}
which are consistent with \eqref{Transport}.
Since the gauge field in the Newton-Cartan theory
\begin{align}
 [B_t] &= 2z - 2 \ , &
 [B_i] &= z-1 \ ,
\end{align}
the scaling dimension of mass $m$ is given by
\begin{equation}
 [m] = - z + 2 - \frac{\theta}{d-1} \ .
\end{equation}


\newpage






\addcontentsline{toc}{section}{References}
\begin{thebibliography}{99}

  \bibitem{km1}
  E.~Kiritsis and Y.~Matsuo,
  {\em ``Charge-hyperscaling violating Lifshitz hydrodynamics from black-holes,''}
  JHEP {\bf 1512} (2015) 076  doi:10.1007/JHEP12(2015)076
  \hri{1508.02494}{[hep-th]}.

\bibitem{Maldacena:1997re}
  J.~M.~Maldacena,
 {\em  ``The large N limit of superconformal field theories and supergravity,''}
  Adv.\ Theor.\ Math.\ Phys.\  {\bf 2} (1998) 231
  [Int.\ J.\ Theor.\ Phys.\  {\bf 38} (1999) 1113]
  \hre{hep-th}{9711200}.

\bibitem{Gubser:1998bc}
  S.~S.~Gubser, I.~R.~Klebanov and A.~M.~Polyakov,
  {\em ``Gauge theory correlators from non-critical string theory,''}
  Phys.\ Lett.\ B {\bf 428} (1998) 105
  \hre{hep-th}{9802109}.

\bibitem{Witten:1998qj}
  E.~Witten,
  {\em ``Anti-de Sitter space and holography,''}
  Adv.\ Theor.\ Math.\ Phys.\  {\bf 2} (1998) 253
  \hre{hep-th}{9802150}.

\bibitem{Witten:1998zw}
  E.~Witten,
  {\em ``Anti-de Sitter space, thermal phase transition, and confinement in  gauge theories,''}
  Adv.\ Theor.\ Math.\ Phys.\  {\bf 2} (1998) 505
  \hre{hep-th}{9803131}.


\bibitem{Son:2002sd}
  D.~T.~Son and A.~O.~Starinets,
  {\em ``Minkowski space correlators in AdS / CFT correspondence: Recipe and applications,''}
  JHEP {\bf 0209} (2002) 042
  \hre{hep-th}{0205051}.

\bibitem{Policastro:2002se}
  G.~Policastro, D.~T.~Son and A.~O.~Starinets,
  {\em ``From AdS / CFT correspondence to hydrodynamics,''}
  JHEP {\bf 0209} (2002) 043
  \hre{hep-th}{0205052}.

\bibitem{Policastro:2002tn}
  G.~Policastro, D.~T.~Son and A.~O.~Starinets,
  {\em ``From AdS / CFT correspondence to hydrodynamics. 2. Sound waves,''}
  JHEP {\bf 0212} (2002) 054
  \hre{hep-th}{0210220}.

\bibitem{Baier:2007ix}
  R.~Baier, P.~Romatschke, D.~T.~Son, A.~O.~Starinets and M.~A.~Stephanov,
  {\em ``Relativistic viscous hydrodynamics, conformal invariance, and holography,''}
  JHEP {\bf 0804} (2008) 100
  \hri{0712.2451}{[hep-th]}.

\bibitem{Bhattacharyya:2008jc}
  S.~Bhattacharyya, V.~E.~Hubeny, S.~Minwalla and M.~Rangamani,
  {\em ``Nonlinear Fluid Dynamics from Gravity,''}
  JHEP {\bf 0802} (2008) 045
  \hri{0712.2456}{[hep-th]}.

\bibitem{Bhattacharyya:2008xc}
  S.~Bhattacharyya, V.~E.~Hubeny, R.~Loganayagam, G.~Mandal, S.~Minwalla, T.~Morita, M.~Rangamani and H.~S.~Reall,
  {\em ``Local Fluid Dynamical Entropy from Gravity,''}
  JHEP {\bf 0806} (2008) 055
  \hri{0803.2526}{[hep-th]}.



\bibitem{Son:2008ye}
  D.~T.~Son,
  {\em ``Toward an AdS/cold atoms correspondence: A Geometric realization of the Schrodinger symmetry,''}
  Phys.\ Rev.\ D {\bf 78} (2008) 046003
  \hri{0804.3972}{[hep-th]}.

\bibitem{Balasubramanian:2008dm}
  K.~Balasubramanian and J.~McGreevy,
  {\em ``Gravity duals for non-relativistic CFTs,''}
  Phys.\ Rev.\ Lett.\  {\bf 101} (2008) 061601
  \hri{0804.4053}{[hep-th]}.

\bibitem{kachru}
  S.~Kachru, X.~Liu and M.~Mulligan,
  {\em ``Gravity duals of Lifshitz-like fixed points,''}
  Phys.\ Rev.\ D {\bf 78} (2008) 106005
  \hri{0808.1725}{[hep-th]}.

\bibitem{taylor}
  M.~Taylor,
  {\em ``Non-relativistic holography,''}
  \hri{0812.0530}{[hep-th]}.

\bibitem{Guica:2010sw}
  M.~Guica, K.~Skenderis, M.~Taylor and B.~C.~van Rees,
  {\em ``Holography for Schrodinger backgrounds,''}
  JHEP {\bf 1102} (2011) 056
  \hri{1008.1991}{[hep-th]}.

  \bibitem{cgkkm}
  C.~Charmousis, B.~Gouteraux, B.~S.~Kim, E.~Kiritsis and R.~Meyer,
  {\em ``Effective Holographic Theories for low-temperature condensed matter systems,''}
  JHEP {\bf 1011} (2010) 151
  \hri{1005.4690}{[hep-th]}.

\bibitem{gk1}
  B.~Gouteraux and E.~Kiritsis,
  {\em ``Generalized Holographic Quantum Criticality at Finite Density,''}
  JHEP {\bf 1112} (2011) 036
  \hri{1107.2116}{[hep-th]}.

\bibitem{gk2}
  B.~Gouteraux and E.~Kiritsis,
  {\em ``Quantum critical lines in holographic phases with (un)broken symmetry,''}
  JHEP {\bf 1304} (2013) 053
  \hri{1212.2625}{[hep-th]}.

  \bibitem{Marika}
  M.~Taylor,
  {\em ``Lifshitz holography,''}
  Class.\ Quant.\ Grav.\  {\bf 33} (2016) no.3,  033001
  \hri{1512.03554}{[hep-th]}.




\bibitem{Eagles:1969zz}
  D.~M.~Eagles,
  {\em ``Possible Pairing without Superconductivity at Low Carrier Concentrations in Bulk and Thin-Film Superconducting Semiconductors,''}
  Phys.\ Rev.\  {\bf 186} (1969) 456.

\bibitem{Nozieres:1985zz}
  P.~Nozieres and S.~Schmitt-Rink,
  {\em ``Bose condensation in an attractive fermion gas: From weak to strong coupling superconductivity,''}
  J.\ Low.\ Temp.\ Phys.\  {\bf 59} (1985) 195.

\bibitem{O'Hara:2002zz}
  K.~M.~O'Hara, S.~L.~Hemmer, M.~E.~Gehm, S.~R.~Granade and J.~E.~Thomas,
  {\em ``Observation of a Strongly Interacting Degenerate Fermi Gas of Atoms,''}
  Science {\bf 298} (2002) 2179
  \hre{cond-mat}{0212463} [cond-mat.supr-con].

\bibitem{Regal:2004zza}
  C.~A.~Regal, M.~Greiner and D.~S.~Jin,
  {\em ``Observation of Resonance Condensation of Fermionic Atom Pairs,''}
  Phys.\ Rev.\ Lett.\  {\bf 92} (2004) 040403.

\bibitem{Bartenstein:2004zza}
  M.~Bartenstein, A.~Altmeyer, S.~Riedl, S.~Jochim, C.~Chin, J.~H.~Denschlag and R.~Grimm,
  {\em ``Crossover from a Molecular Bose-Einstein Condensate to a Degenerate Fermi Gas,''}
  Phys.\ Rev.\ Lett.\  {\bf 92} (2004) 120401.

  \bibitem{Patel:2016ymd}
  A.~A.~Patel, A.~Eberlein and S.~Sachdev,
  {\em ``Shear viscosity at the Ising-nematic quantum critical point in two dimensional metals,''}
  \hri{1607.03894}{[cond-mat.str-el]}.




\bibitem{Hornreich:1975zz}
  R.~M.~Hornreich, M.~Luban and S.~Shtrikman,
  {\em ``Critical Behavior at the Onset of k-Space Instability on the lamda Line,''}
  Phys.\ Rev.\ Lett.\  {\bf 35} (1975) 1678.

\bibitem{Mehen:1999nd}
  T.~Mehen, I.~W.~Stewart and M.~B.~Wise,
  {\em ``Conformal invariance for nonrelativistic field theory,''}
  Phys.\ Lett.\ B {\bf 474} (2000) 145
  \hre{hep-th}{9910025}.

\bibitem{Ardonne:2003wa}
  E.~Ardonne, P.~Fendley and E.~Fradkin,
  {\em ``Topological order and conformal quantum critical points,''}
  Annals Phys.\  {\bf 310} (2004) 493
  \hre{cond-mat}{0311466}.


\bibitem{Herzog:2008wg}
  C.~P.~Herzog, M.~Rangamani and S.~F.~Ross,
  {\em ``Heating up Galilean holography,''}
  JHEP {\bf 0811} (2008) 080
  \hri{0807.1099}{[hep-th]}.

\bibitem{Rangamani:2008gi}
  M.~Rangamani, S.~F.~Ross, D.~T.~Son and E.~G.~Thompson,
  {\em ``Conformal non-relativistic hydrodynamics from gravity,''}
  JHEP {\bf 0901} (2009) 075
  \hri{0811.2049}{[hep-th]}.

  \bibitem{Bra}
 D.~K.~Brattan,
  {\em ``Charged, conformal non-relativistic hydrodynamics,''}
  JHEP {\bf 1010} (2010) 015
  \hri{1003.0797}{[hep-th]}.


\bibitem{Hoyos:2013qna}
  C.~Hoyos, B.~S.~Kim and Y.~Oz,
  {\em ``Lifshitz Field Theories at Non-Zero Temperature, Hydrodynamics and Gravity,''}
  JHEP {\bf 1403} (2014) 029
  \hri{1309.6794}{[hep-th]}.

\bibitem{Hoyos:2013cba}
  C.~Hoyos, B.~S.~Kim and Y.~Oz,
  {\em ``Bulk Viscosity in Holographic Lifshitz Hydrodynamics,''}
  JHEP {\bf 1403} (2014) 050
  \hri{1312.6380}{[hep-th]}.



  \bibitem{fu}
  C.~Cao, E.~Elliott, J.~Joseph, H.~Wu, J.~Petricka, T.~Schafer and J.~E.~Thomas,
  {\em ``Universal Quantum Viscosity in a Unitary Fermi Gas,''}
  Science {\bf 331} (2011) 58
  \hri{1007.2625}{[cond-mat.quant-gas]}.

\bibitem{zaa}
  D.~Forcella, J.~Zaanen, D.~Valentinis and D.~van der Marel,
  {\em ``Electromagnetic properties of viscous charged fluids,''}
  Phys.\ Rev.\ B {\bf 90} (2014) 3,  035143
  \hri{1406.1356}{[cond-mat.str-el]}.

\bibitem{gra}
  M.~M\"uller and S.~Sachdev,
  {\em ``Collective cyclotron motion of the relativistic plasma in graphene,''}
  Phys.\ Rev.\ B {\bf 78} (2008) 115419
  \hri{0801.2970}{[cond-mat.str-el]};\\
  L. Fritz, J. Schmalian, M. M\"uller and S Sachdev,
  {\em ``Quantum critical transport in clean graphene, "}
   	Phys.\ Rev.\ B{\bf 78}, (2008) 085416
\hri{0802.4289}{[cond-mat.mes-hall]};\\
M. Mueller, J. Schmalian, L. Fritz
{\em ``Graphene - a nearly perfect fluid, "}
 	Phys. Rev. Lett. {\bf 103}, (2009) 025301
\hri{0903.4178}{[cond-mat.mes-hall]}



\bibitem{g1}
 D. A. Bandurin, I. Torre, R. K. Kumar, M. Ben Shalom,
A. Tomadin, A. Principi, G. H. Auton, E. Khestanova,
K. S. Novoselov, I. V. Grigorieva, L. A. Ponomarenko,
A. K. Geim, and M. Polini,
{\em  ``Negative local resistance
caused by viscous electron back
ow in graphene"}, Science 351, 1055 (2016), \hri{1509.04165}{[cond-mat.str-el]}.

\bibitem{g2} J. Crossno, J. K. Shi, K. Wang, X. Liu, A. Harzheim,
A. Lucas, S. Sachdev, P. Kim, T. Taniguchi, K. Watanabe,
T. A. Ohki, and K. C. Fong,
{\em ``Observation of
the Dirac fluid and the breakdown of the Wiedemann-
Franz law in graphene,"}
Science 351, 1058 (2016),
\hri{1509.04713}{[cond-mat.mes-hall]}.


  \bibitem{g3} P. J. W. Moll, P. Kushwaha, N. Nandi, B. Schmidt,
and A. P. Mackenzie, {\em ``Evidence for hydrodynamic electron flow in PdCoO2"}, Science 351, 1061 (2016),
\hri{1509.05691}{[cond-mat.str-el]}.


\bibitem{g4}  I. Torre,  A. Tomadin,  A. K. Geim,    and M. Polini,
{\em Nonlocal transport and the hydrodynamic shear viscosity in graphene"}, Phys. Rev. B 92, 165433(2015),
 \hri{1508.00363}{[cond-mat.mes-hall]}.

\bibitem{g5} L. Levitov and G. Falkovich, {\em ``Electron viscosity, current vortices and negative nonlocal resistance in graphene"},
     Nature Physics 12, 672 (2016), \hri{1508.00836}{[cond-mat.mes-hall]}.

\bibitem{g6} A.  Lucas,  J.  Crossno,  K.  C.  Fong,  P.  Kim,    and
S. Sachdev, {\em ``Transport in inhomogeneous quantum criti- cal fluids and in the Dirac fluid in graphene"},  Phys. Rev.  {\bf B}93, 075426 (2016), \hri{1510.01738}{[cond-mat.str-el]}.








  \bibitem{hartong-obers}
  J.~Gath, J.~Hartong, R.~Monteiro and N.~A.~Obers,
  {\em ``Holographic Models for Theories with Hyperscaling Violation,''}
  JHEP {\bf 1304} (2013) 159
  \hri{1212.3263}{[hep-th]}.


    \bibitem{G1}
  B.~Gouteraux,
  {\em ``Universal scaling properties of extremal cohesive holographic phases,''}
  JHEP {\bf 1401} (2014) 080
  \hri{1308.2084}{[hep-th]}.


  \bibitem{karch}
  A.~Karch,
  {\em ``Conductivities for Hyperscaling Violating Geometries,''}
  JHEP {\bf 1406} (2014) 140
  \hri{1405.2926}{[hep-th]}.


\bibitem{sh}
  L.~Huijse, S.~Sachdev and B.~Swingle,
  {\em ``Hidden Fermi surfaces in compressible states of gauge-gravity duality,''}
  Phys.\ Rev.\ B {\bf 85} (2012) 035121
  \hri{1112.0573}{[cond-mat.str-el]}.


%
\bibitem{Son}
  D.~T.~Son,
  {\em ``Newton-Cartan Geometry and the Quantum Hall Effect,''}
  \hri{1306.0638}{[cond-mat.mes-hall]}.



\bibitem{Christensen:2013lma}
  M.~H.~Christensen, J.~Hartong, N.~A.~Obers and B.~Rollier,
  {\em ``Torsional Newton-Cartan Geometry and Lifshitz Holography,''}
  Phys.\ Rev.\ D {\bf 89} (2014) 061901
  \hri{1311.4794}{[hep-th]}.

\bibitem{Christensen:2013rfa}
  M.~H.~Christensen, J.~Hartong, N.~A.~Obers and B.~Rollier,
  {\em ``Boundary Stress-Energy Tensor and Newton-Cartan Geometry in Lifshitz Holography,''}
  JHEP {\bf 1401} (2014) 057
  \hri{1311.6471}{[hep-th]}.


  \bibitem{BHR}
  E.~A.~Bergshoeff, J.~Hartong and J.~Rosseel,
  {\em ``Torsional Newton-Cartan geometry and the Schrodinger algebra,''}
  Class.\ Quant.\ Grav.\  {\bf 32} (2015) 13,  135017
  \hri{1409.5555}{[hep-th]}.

\bibitem{Hartong:2014oma}
  J.~Hartong, E.~Kiritsis and N.~A.~Obers,
  {\em ``Lifshitz space-times for Schr\"odinger holography,''}
  Phys.\ Lett.\ B {\bf 746} (2015) 318
  \hri{1409.1519}{[hep-th]}.

\bibitem{Hartong:2014pma}
  J.~Hartong, E.~Kiritsis and N.~A.~Obers,
  {\em ``Schroedinger Invariance from Lifshitz Isometries in Holography and Field Theory,''}
  \hri{1409.1522}{[hep-th]}.

\bibitem{Hartong:2015wxa}
  J.~Hartong, E.~Kiritsis and N.~A.~Obers,
  {\em ``Field Theory on Newton-Cartan Backgrounds and Symmetries of the Lifshitz Vacuum,''}
  \hri{1502.00228}{[hep-th]}.



\bibitem{RS}
  S.~F.~Ross and O.~Saremi,
  {\em ``Holographic stress tensor for non-relativistic theories,''}
  JHEP {\bf 0909} (2009) 009
  \hri{0907.1846}{[hep-th]}.

\bibitem{CP}
  W.~Chemissany and I.~Papadimitriou,
  {\em ``Lifshitz holography: The whole shebang,''}
  JHEP {\bf 1501} (2015) 052;
  \hri{1408.0795}{[hep-th]};\\
  {\em ``Generalized dilatation operator method for non-relativistic holography,''}
  Phys.\ Lett.\ B {\bf 737} (2014) 272;
  \hri{1405.3965}{[hep-th]}.

  \bibitem{GJSV}
  U.~Gursoy, A.~Jansen, W.~Sybesma and S.~Vandoren,
  {\em ``Holographic Equilibration of Nonrelativistic Plasmas,''}
  \hri{1602.01375}{[hep-th]}.

\bibitem{HO}
  J.~Hartong, N.~A.~Obers and M.~Sanchioni,
  {\em ``Lifshitz Hydrodynamics from Lifshitz Black Branes with Linear Momentum,''}
  \hri{1606.09543}{[hep-th]}.

  \bibitem{kk}
  E.~Kiritsis,
  {\em ``Supergravity, D-brane probes and thermal superYang-Mills: A Comparison,''}
  JHEP {\bf 9910} (1999) 010
\hre{hep-th}{9906206}.

  \bibitem{tw}
  S.~S.~Gubser,
  {\em ``Time warps,''}
  JHEP {\bf 1001} (2010) 020
  \hri{0812.5107}{[hep-th]}.

  \bibitem{gn}
  S.~S.~Gubser and A.~Nellore,
  ``Ground states of holographic superconductors,''
  Phys.\ Rev.\ D {\bf 80} (2009) 105007
  doi:10.1103/PhysRevD.80.105007
  \hri{0908.1972}{[hep-th]}.

  \bibitem{mukho}
  S.~Kuperstein and A.~Mukhopadhyay,
  {\em ``The unconditional RG flow of the relativistic holographic fluid,''}
  JHEP {\bf 1111} (2011) 130
  \hri{1105.4530}{[hep-th]};\\
  S.~Kuperstein and A.~Mukhopadhyay,
  {\em ``Spacetime emergence via holographic RG flow from incompressible Navier-Stokes at the horizon,''}
  JHEP {\bf 1311} (2013) 086
  \hri{1307.1367}{[hep-th]};\\
  N.~Behr, S.~Kuperstein and A.~Mukhopadhyay,
  {\em `Holography as a highly efficient renormalization group flow. I. Rephrasing gravity,''}
  Phys.\ Rev.\ D {\bf 94} (2016) no.2,  026001
  \hri{1502.06619}{[hep-th]}.

\bibitem{Ross:2011gu}
  S.~F.~Ross,
  {\em ``Holography for asymptotically locally Lifshitz spacetimes,''}
  Class.\ Quant.\ Grav.\  {\bf 28} (2011) 215019
  \hri{1107.4451}{[hep-th]}.




\bibitem{Kovtun:2004de}
  P.~Kovtun, D.~T.~Son and A.~O.~Starinets,
  {\em ``Viscosity in strongly interacting quantum field theories from black hole physics,''}
  Phys.\ Rev.\ Lett.\  {\bf 94} (2005) 111601
  \hre{hep-th}{0405231}.




\bibitem{Jensen:2014ama}
  K.~Jensen,
  {\em ``Aspects of hot Galilean field theory,''}
  JHEP {\bf 1504} (2015) 123
  \hri{1411.7024}{[hep-th]}.

\bibitem{Jensen:2014aia}
  K.~Jensen,
  {\em ``On the coupling of Galilean-invariant field theories to curved spacetime,''}
  \hri{1408.6855}{[hep-th]}.



\bibitem{Kanitscheider:2009as}
  I.~Kanitscheider and K.~Skenderis,
  {\em ``Universal hydrodynamics of non-conformal branes,''}
  JHEP {\bf 0904} (2009) 062  doi:10.1088/1126-6708/2009/04/062
\hri{0901.1487}{[hep-th]}.

\bibitem{gs}
  B.~Gouteraux, J.~Smolic, M.~Smolic, K.~Skenderis and M.~Taylor,
  {\em ``Holography for Einstein-Maxwell-dilaton theories from generalized dimensional reduction,''}
  JHEP {\bf 1201} (2012) 089
\hri{1110.2320}{[hep-th]}.


\bibitem{ho3}
  J.~Hartong and N.~A.~Obers,
  {\em ``Horava-Lifshitz gravity from dynamical Newton-Cartan geometry,''}
  JHEP {\bf 1507} (2015) 155;
  \hri{1504.07461}{[hep-th]}.

  \bibitem{nakayama}
  Y.~Nakayama,
  {\em ``Intrinsic ambiguity in second order viscosity parameters in relativistic hydrodynamics,''}
  Int.\ J.\ Mod.\ Phys.\ A {\bf 27} (2012) 1250125
  doi:10.1142/S0217751X12501254
  \hri{1206.2421}{[hep-th]}.

\end{thebibliography}
\end{document}